%% file: main.tex
\documentclass[journal,11pt,onecolumn,draftclsnofoot]{article}

\usepackage{amsmath,amssymb}
\usepackage{graphicx}
\usepackage{subcaption}
\usepackage{booktabs}
\usepackage{siunitx}
\usepackage[margin=1in]{geometry}
\usepackage{hyperref} 

\usepackage{amsmath}
\usepackage{amssymb}
\usepackage{float}
\usepackage{braket}
\usepackage{xcolor}
\usepackage{mathrsfs}
\usepackage{graphicx}
\usepackage{mathrsfs}  
\usepackage{siunitx}
\usepackage{hyperref}
\usepackage{subcaption}
\usepackage{lineno}
\usepackage{subcaption}
\usepackage{setspace}
\usepackage[margin=1in]{geometry} 
\usepackage{booktabs}
\usepackage{multirow}
\usepackage{threeparttable}
\usepackage{rotating}
\usepackage{soul}
\usepackage{xr}
\usepackage{empheq}

\usepackage{textcomp}

\usepackage{authblk}
\usepackage{hyperref}

\DeclareCaptionLabelFormat{bold}{\textbf{#2}}
\begin{document}

\title{\textbf{Microwave dielectric properties of LiNbO$_{\mathbf{3}}$ and AlN at millikelvin temperatures and single-photon power}}



\author[1,2]{Alessandro Reineri}
\author[1]{Francesco Crisa}
\author[1]{Maithlee Shinde}
\author[1]{Daniel Bafia}
\author[1]{Akshay Murthy}
\author[1]{Changqing Wang}
\author[1]{Tanay Roy}
\author[1]{Alexander Romanenko}
\author[1,2]{John Zasadzinski}
\author[1]{Anna Grassellino}
\author[1,3,*]{Silvia Zorzetti}

\affil[1]{Fermi National Accelerator Laboratory, Batavia, IL, USA}
\affil[2]{Illinois Institute of Technology, Chicago, IL, USA}
\affil[3]{Northwestern University, Evanston, IL, USA}
\affil[*]{Corresponding author: zorzetti@ieee.org}
\date{}  

\maketitle
\thispagestyle{plain}
\pagestyle{plain}

\section*{\textbf{Abstract}}
\input{abstract}

\section*{\textbf{Introduction}}
\input{introduction}
\section*{\textbf{Results}}
\input{results}

\section*{\textbf{Methods}}
\input{methods}

\section*{\textbf{Discussion}}
\input{conclusions}

\section*{\bf Funding}

This manuscript has been authored by the Fermi Forward Discovery Group, LLC under Contract No. 89243024CSC000002 with the U.S. Department of Energy, Office of Science, Office of High Energy Physics. This material is based upon work supported by the U.S. Department of Energy, Office of Science, Early Career Research Program.
This work was partially supported by the U.S. Department of Energy, Office of Science, National Quantum Information Science Research Centers, Superconducting Quantum Materials and Systems Center (SQMS), under Contract No. 89243024CSC000002. 

\section*{\bf Acknowledgment}
The SQMS Center supported access to facilities, equipment, and measurement of electro-optic crystals in superconducting cavities. The authors thank Dr. Vyacheslav Yakovlev for insightful discussions and feedback on this work, and Dr. David van Zanten for support with testbeds and operations.

\section*{\textbf{Author information}}

\subsection*{Author Contributions}
S.Z. conceived the project, coordinated the research activities, and assisted in drafting and editing the manuscript. A.R. assembled the cavities, conducted microwave simulations for optimal sample positioning, and performed power- and temperature-dependent measurements. A.R. and F.C. performed the data analysis. M.S. and A.M. performed room temperature XPS and ToF-SIMS material characterization and analyses. C.W. contributed to the initial conception of the experiment and provided use-cases perspectives related to quantum transduction. D.B. T.R., A.G. A.Rom., and J.Z. contributed to the validation and interpretation of the results and to the final version of the manuscript. All authors reviewed the manuscript. 
\subsection*{Corresponding authors}
Correspondence to Silvia Zorzetti, email: \href{zorzetti@ieee.org}{zorzetti@ieee.org}.

\section*{\textbf{Declarations}}

\subsection*{\bf Data Availability}
The numerical data generated in this work is available from the authors upon reasonable request. 

\subsection*{\bf Code Availability}
The code generated in this work is available from the authors upon reasonable request.

\subsection*{\bf Competing Interests}
The authors declare that they have no competing interests. 

\subsection*{\bf Ethics approval and consent to participate}
Not applicable.

\bibliographystyle{naturemag}
\bibliography{references}

\end{document}


\title{\textbf{Supplementary Information: Microwave dielectric properties of LiNbO$_{\mathbf{3}}$ and AlN at millikelvin temperatures and single-photon power}}

\author[1,2]{Alessandro Reineri}
\author[1]{Francesco Crisa}
\author[1]{Maithlee Shinde}
\author[1]{Daniel Bafia}
\author[1]{Akshay Murthy}
\author[1]{Changqing Wang}
\author[1]{Tanay Roy}
\author[1]{Alexander Romanenko}
\author[1,2]{John Zasadzinski}
\author[1]{Anna Grassellino}
\author[1,3, *]{Silvia Zorzetti}

\affil[1]{Fermi National Accelerator Laboratory, Batavia, IL, USA}
\affil[2]{Illinois Institute of Technology, Chicago, IL, USA}
\affil[3]{Northwestern University, Evanston, IL, USA}
\affil[*]{Corresponding author: zorzetti@ieee.org}
\date{}

\maketitle
\thispagestyle{plain}
\pagestyle{plain}

In this supplementary material, we discuss in more detail the cavity design, the preliminary finite-element electromagnetic (FEM) simulations performed, and the samples and cavity preparation for the experiment. Then, we expand on the circle-fit routine implemented to analyze the reflection data, highlighting its optimal working regime and a strategy to mitigate possible amplifier gain compression. Subsequently, we give a qualitative explanation for the base-temperature high-power $Q_{0}$ behavior observed in the crystals TM$_{010}$ resonances. Finally, we report the observed temperature-dependent relative frequency shift during the temperature sweeps for all resonances and qualitatively describe the data behavior.

\begin{figure}[htbp!]
    \centering
    \begin{subfigure}[t]{0.48\linewidth}
        \centering
        \subcaption{\raisebox{-20ex}[0pt][0pt]{\hspace{130pt}}}
        \includegraphics[scale = 0.38]{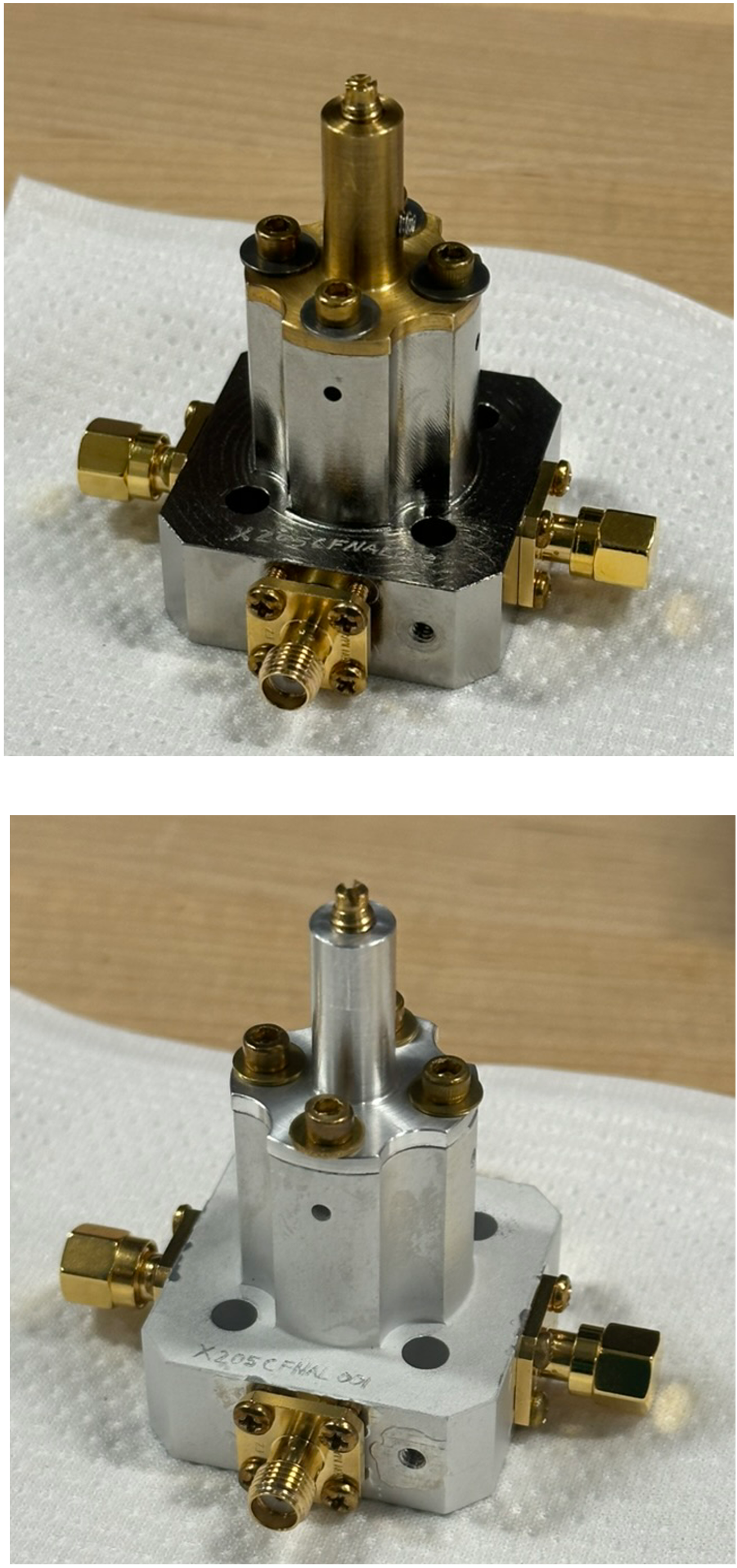}
        \label{fig:CavityPhoto}
    \end{subfigure}
    \quad
    \begin{subfigure}[t]{0.48\linewidth}
        \centering
        \subcaption{\raisebox{66ex}[0pt][0pt]{\hspace{170pt}}}
        \includegraphics[scale = 0.47]{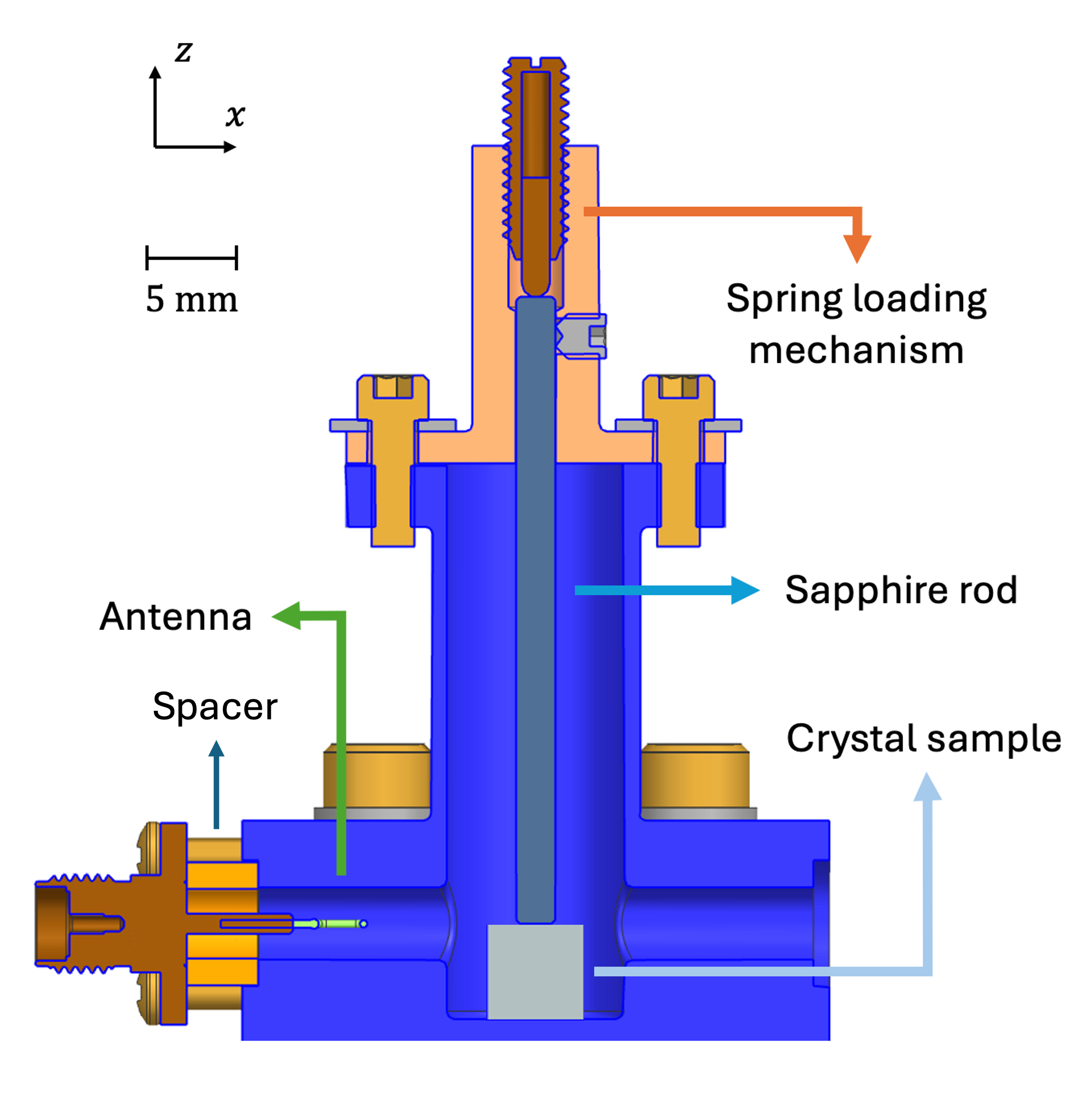}
        \label{fig:CavityCADsection}
    \end{subfigure}
    \caption{(a) Picture of the two cavities used to test the LiNbO$_{3}$ and the AlN samples, the upper one made of niobium and the lower one of aluminum. (b) Cross-sectional CAD of the cavity design.}
    \label{fig:CavityPicsandCAD}
\end{figure}

\begin{figure}[htbp!]
    \centering
    \begin{subfigure}[b]{0.48\linewidth}
        \centering
        \subcaption{\raisebox{50ex}[0pt][0pt]{\hspace{150pt}}}
        \includegraphics[scale = 0.41]{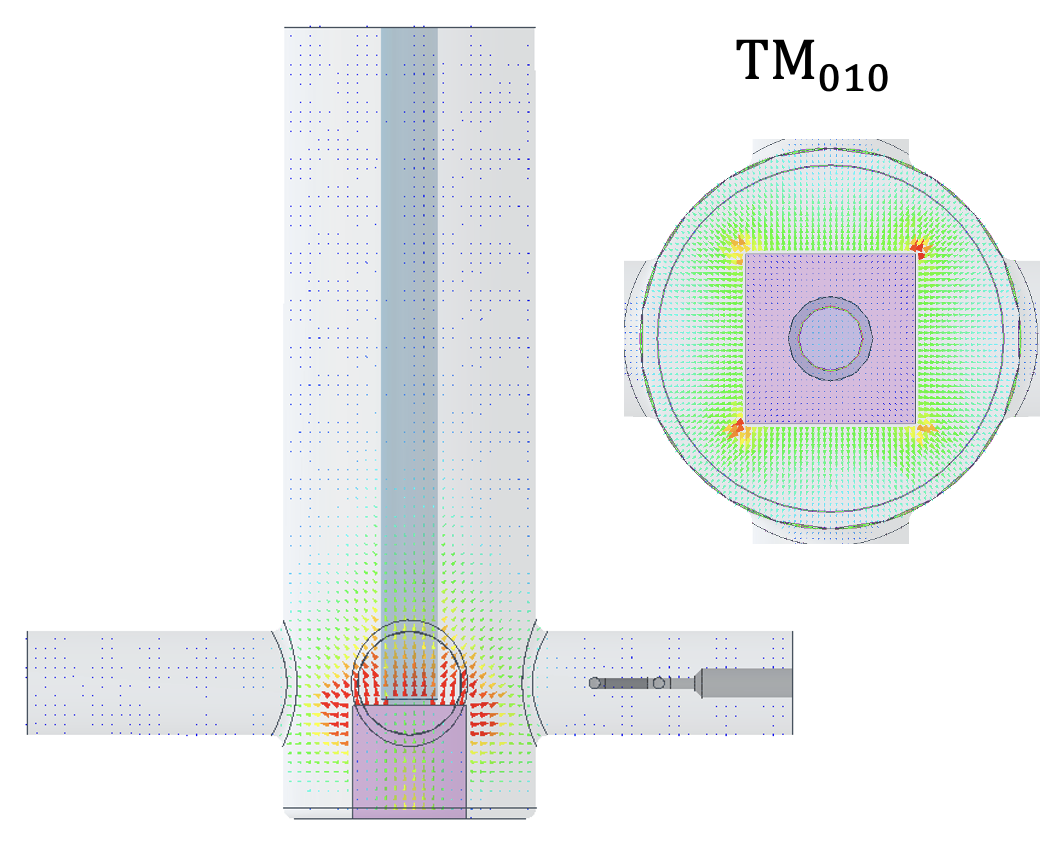}
        \label{fig:TM010distro}
    \end{subfigure}
    \hspace{0.02\linewidth}
    \begin{subfigure}[b]{0.48\linewidth}
        \centering
        \subcaption{\raisebox{50ex}[0pt][0pt]{\hspace{150pt}}}
        \includegraphics[scale = 0.41]{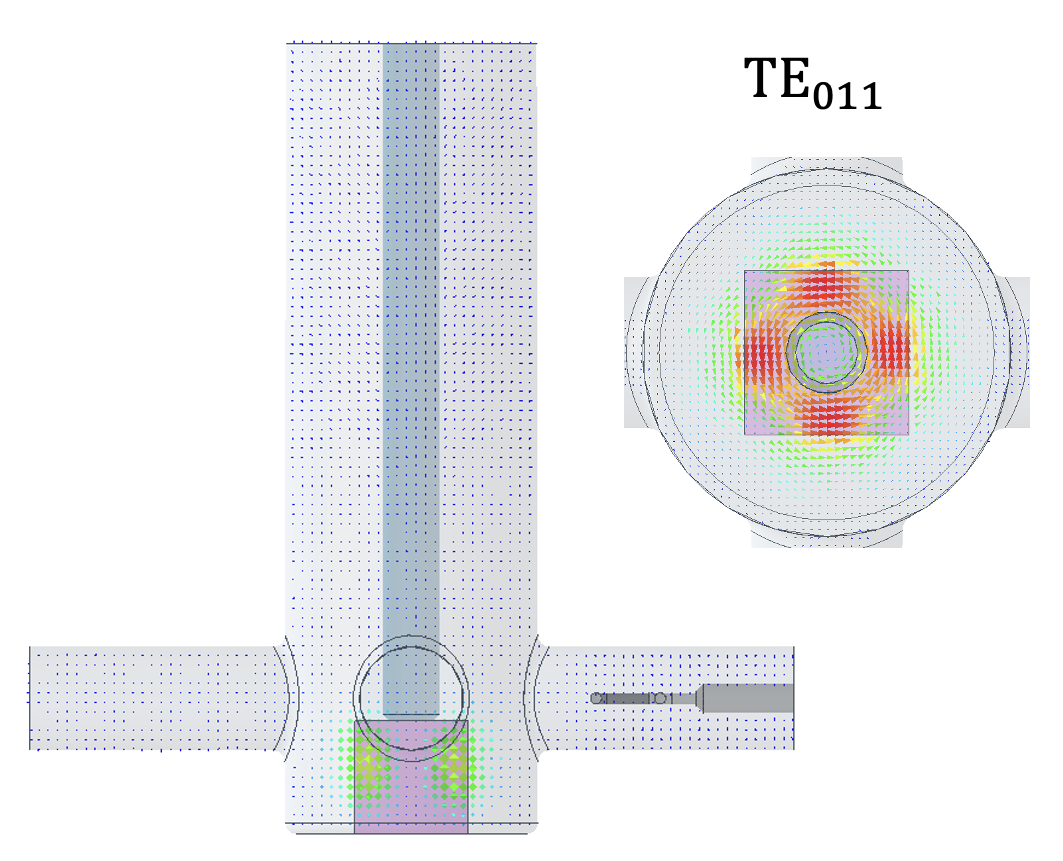}
        \label{fig:TE011distro}
    \end{subfigure}
      \\
   \vspace{0.5cm}
    \caption{Field distribution heatmap of the (a) $\mathrm{TM_{010}}$ and (b) $\mathrm{TE_{011}}$ modes in the cavity in presence of 
       one of the dielectric cubic samples under test.}
        \label{fig:EFieldDistro}
\end{figure}

\supplementarynote{Cavity design, FEM simulations and experiment preparation}
The cavity design is derived from a cylindrical cavity with four ports to position up to four I/O antennas (Suppl.~Fig.~\ref{fig:CavityPicsandCAD}). Two cavities have been manufactured: the first is made of bulk aluminum 6061, and the second is made of high-purity bulk niobium. The shape of the cavity allows us to distinguish between transverse magnetic (TM) and transverse electric (TE) modes by varying the input frequency and exciting the different dielectric directions of the crystals' permittivity\cite{zorzetti2023millikelvin}. The materials under study belong to the class of birefringent crystals, for which the permittivity is modeled by a tensor:
\begin{equation}
    \tag{S1}
    \label{eq:anisoEps}
    \varepsilon = \begin{pmatrix}
        \varepsilon_{\perp} & 0 & 0 \\
        0 & \varepsilon_{\perp} & 0 \\
        0 & 0 & \varepsilon_{\parallel}
    \end{pmatrix},
\end{equation}
with all nonzero entries being complex numbers, i.e., $\varepsilon_{m} = \varepsilon_{m}^{\prime} - i\varepsilon_{m}^{\prime\prime} = \varepsilon_{m}^{\prime}\left(1-i\tan{\delta_{m}}\right)$, $m = \perp, \parallel$, often referred to as the \textit{in-plane} and \textit{out-of-plane} components. The field lines maintain their overall orientation after the sample is inserted in the cavity and a large portion of the field energy is confined inside the samples' volume (Suppl.~Figs.~\ref{fig:TM010distro}~-~\ref{fig:TE011distro}). 
The samples are held in place by a rod made of high-quality crystalline sapphire, with a base-temperature microwave loss tangent of $10^{-7}$. The cavities are processed with Transene aluminum etchant for the aluminum cavity and high-temperature annealing for the niobium one to ensure the bare cavity quality factor $Q_{0}$ is above $10^{8}$ \cite{kudra2020high, romanenko2020three}. In this way, when measuring the quality factor of the resonance modes of the samples, the leading contribution comes from the quality factor of the dielectrics.

\begin{figure}[htbp!]
    \centering
    \includegraphics[width=0.8\linewidth]{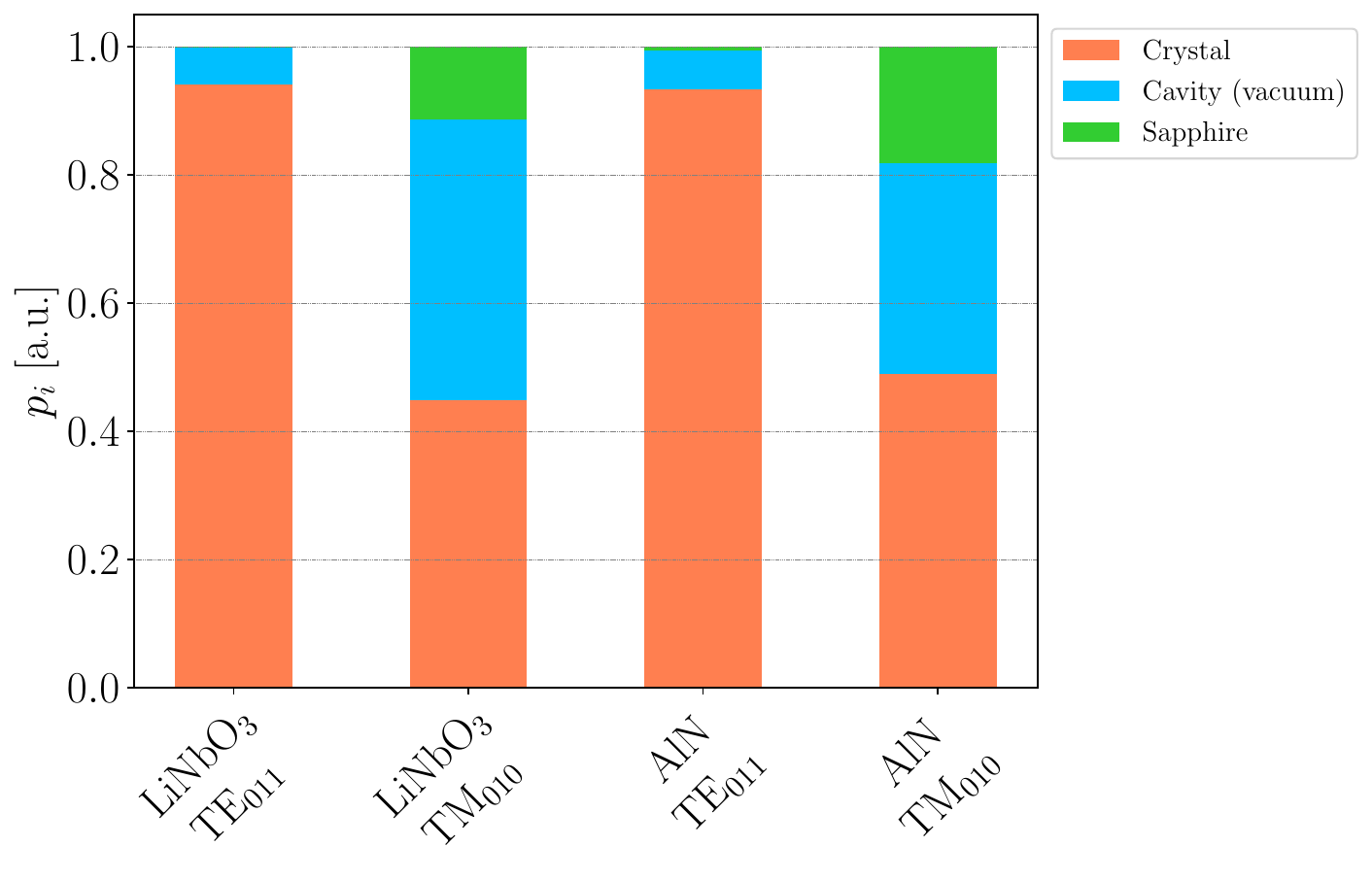}
    \caption{Energy-participation ratio room-temperature simulations for different resonances of the two materials under test. The crystal participation ratio is consistently higher than the sapphire and the cavity vacuum volume ones across all modes, approaching 0.95 for the TE$_{011}$ resonance and 0.45 for the TM$_{010}$. The sapphire participation ratio is either almost null or not exceeding 0.2.}
    \label{fig:PratiosGlob}
\end{figure}

Since the loss tangent of crystalline sapphire is extremely low compared to other dielectric materials for the explored range of frequency at DR base temperature, each resonance quality factor is largely determined by the contribution of the crystal under test. In a more quantitative way, considering energy-participation ratio simulations at room temperature and assuming an \textit{a priori} sample quality factor of 10$^{5}$-10$^6$, the resonance quality factor is equal to the crystal quality factor contribution up to one part in a million. Energy-participation ratio simulations confirm as well that most of the electric field is confined inside the volume of the crystal under test for any given resonance of each material, with the 
second largest contribution coming from the cavity vacuum volume and little to no participation from the sapphire rod (Suppl.~Fig.~\ref{fig:PratiosGlob}). Two cavity materials were used to maximize parallel measurement capabilities per dilution refrigerator cycle.

\begin{figure}[htbp!]
  \centering
  \begin{subfigure}[t]{0.48 \linewidth}
    \centering
    \subcaption{\raisebox{0ex}[0pt][0pt]{\hspace{200pt}}}
    \includegraphics[width=1\linewidth]{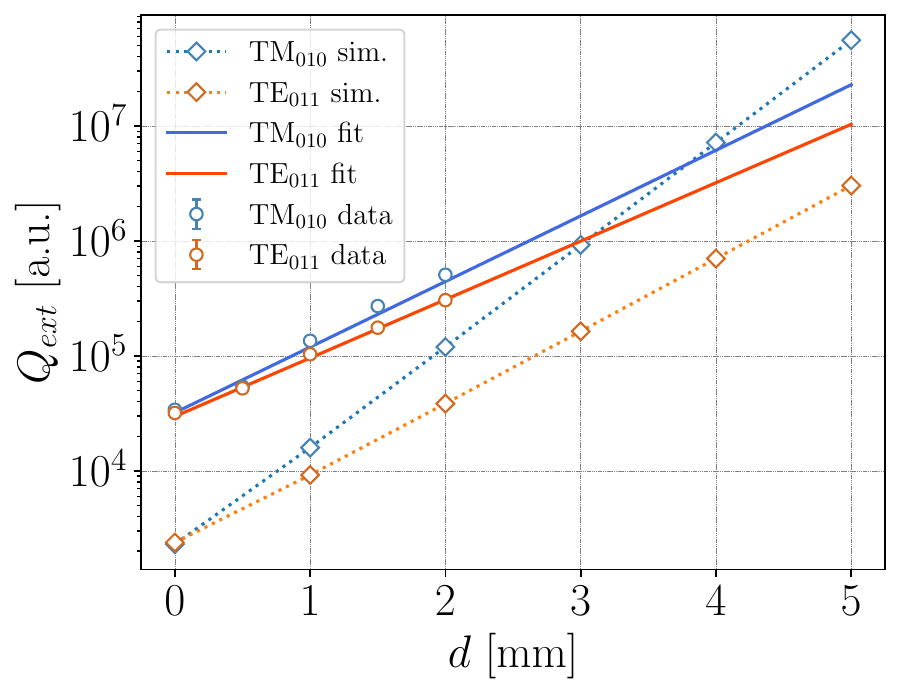}
    \label{fig:QextMeasAndSimVsDRoomTAlN}
  \end{subfigure}
  \quad
  \begin{subfigure}[t]{0.48 \linewidth}
    \centering
    \subcaption{\raisebox{0ex}[0pt][0pt]{\hspace{200pt}}}
    \includegraphics[width=1\linewidth]{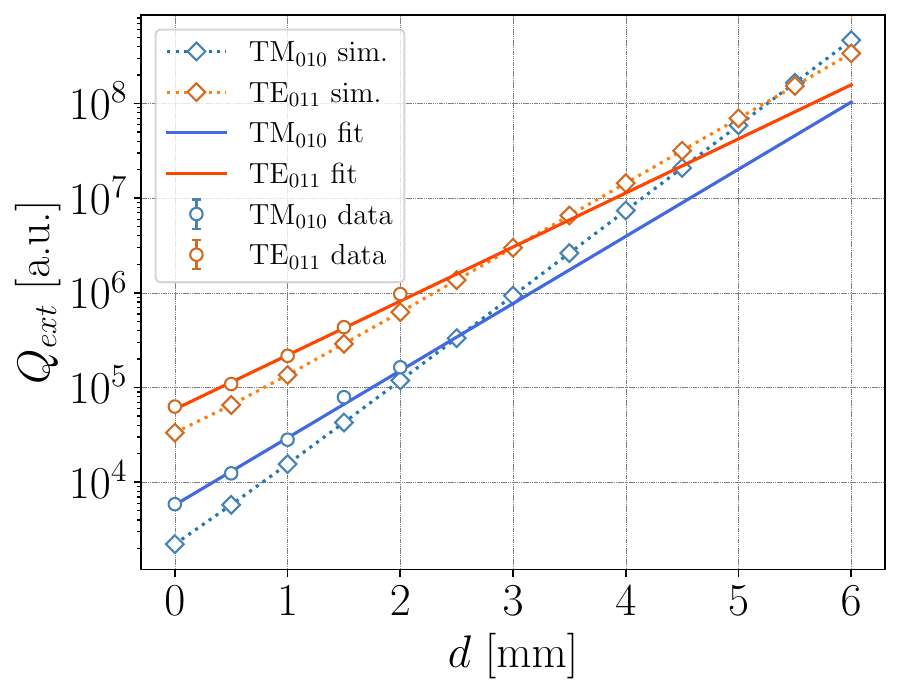}
    \label{fig:QextMeasAndSimVsDRoomTLiNbO3}
  \end{subfigure}
  \caption{Simulations and room temperature measurements of $Q_{ext}$ as a function of the spacer thickness for the cavity with AlN (a) and LiNbO$_{3}$ (b). Larger values of $d$ indicate less antenna protrusion inside the cavity. Measured values are in good agreement with simulations in the case of the LiNbO$_{3}$ sample for both resonances. For the AlN sample, the measured values of $Q_{ext}$ differ from the simulations by one order of magnitude for the deeply protruding antenna positions due to irregularities in sample shape.}
  \label{fig:QextMeasAndSimVsDRoomT}
\end{figure}

A second FEM simulation set was performed to find the optimal length of the I/O antenna to perform resonance parameter extraction with the circle-fit routine for reflection configurations. The external quality factor values of the antenna extracted from simulations are compared to the room-temperature $Q_{ext}$ measurements for the two crystals and for each resonance with different antenna lengths (Suppl.~Fig.~\ref{fig:QextMeasAndSimVsDRoomT}). The antenna protrusion inside the cavity is changed by interposing a brass cylindrical spacer between the antenna holder and the cavity external walls (Suppl.~Fig.~\ref{fig:CavityCADsection}). The optimal antenna length is determined by imposing the condition of critical coupling between antenna and target resonance mode, i.e., $\frac{Q_{0}}{Q_{ext}}=1$ with an \textit{a priori} estimation of $Q_{0} \sim 10^{5}$ \cite{probst2015efficient}. 

In preparation for the base-temperature power sweep measurements, a thorough characterization of the I/O line attenuation was performed with the round-trip method to extract the microwave power delivered to the cavity input port. By sending a tone from the vector network analyzer (VNA) at a frequency detuned by more than three times the target resonance width $\kappa$ and measuring the attenuated signal at the output with a spectrum analyzer, the input power was calculated as the measured output power minus the attenuation along the output branch (see Fig.~1b of the main text). The frequency-dependent attenuation values are reported in Suppl. Tab.~\ref{tab:LineChar}.

\input{Tables/Line_characterization}

\begin{figure}[htbp!]
    \centering
    \begin{subfigure}[t]{0.48\linewidth}
        \centering
        \subcaption{\raisebox{-20ex}[0pt][0pt]{\hspace{130pt}}}
        \includegraphics[width=1\linewidth]{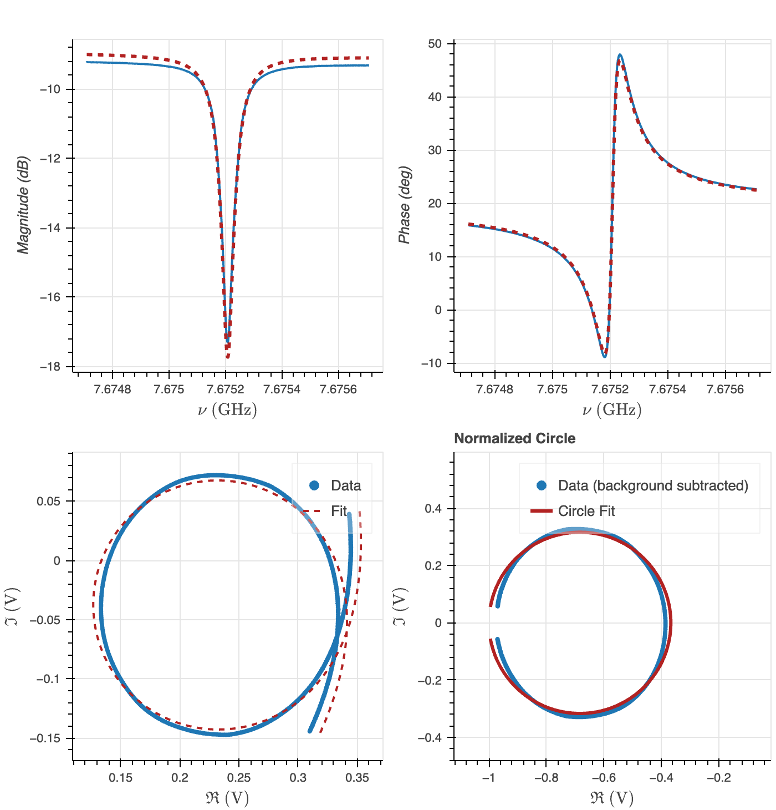}
        \label{fig:CircleFitNoDecompEllipt}
    \end{subfigure}
    \quad
    \begin{subfigure}[t]{0.48\linewidth}
        \centering
        \subcaption{\raisebox{-20ex}[0pt][0pt]{\hspace{130pt}}}
        \includegraphics[width=1\linewidth]{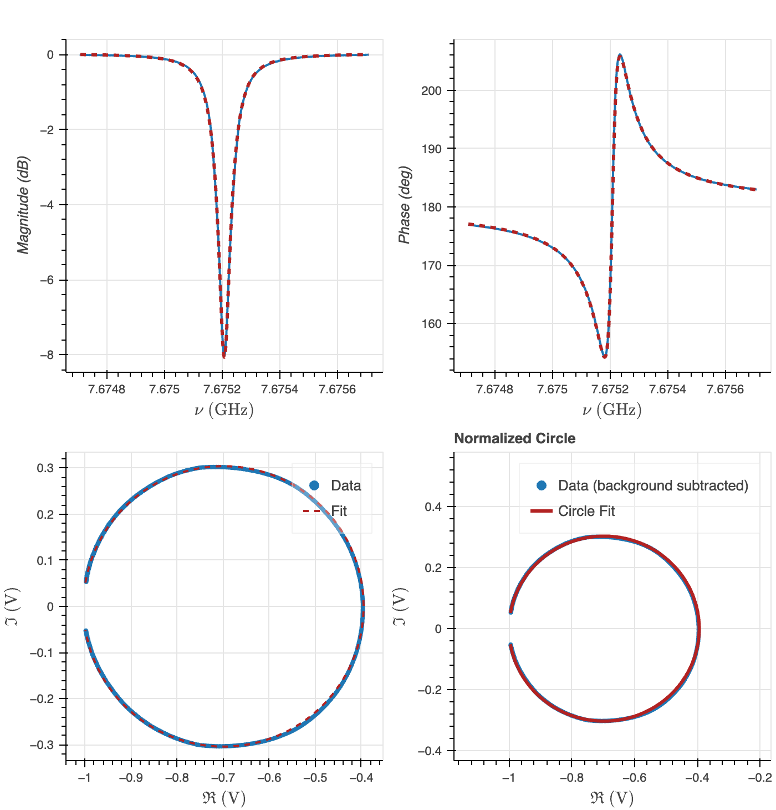}
        \label{fig:CircleFitDecompEllipt}
    \end{subfigure}
    \\
    \begin{subfigure}[t]{0.48\linewidth}
        \centering
        \subcaption{\raisebox{-20ex}[0pt][0pt]{\hspace{130pt}}}
        \includegraphics[width=1\linewidth]{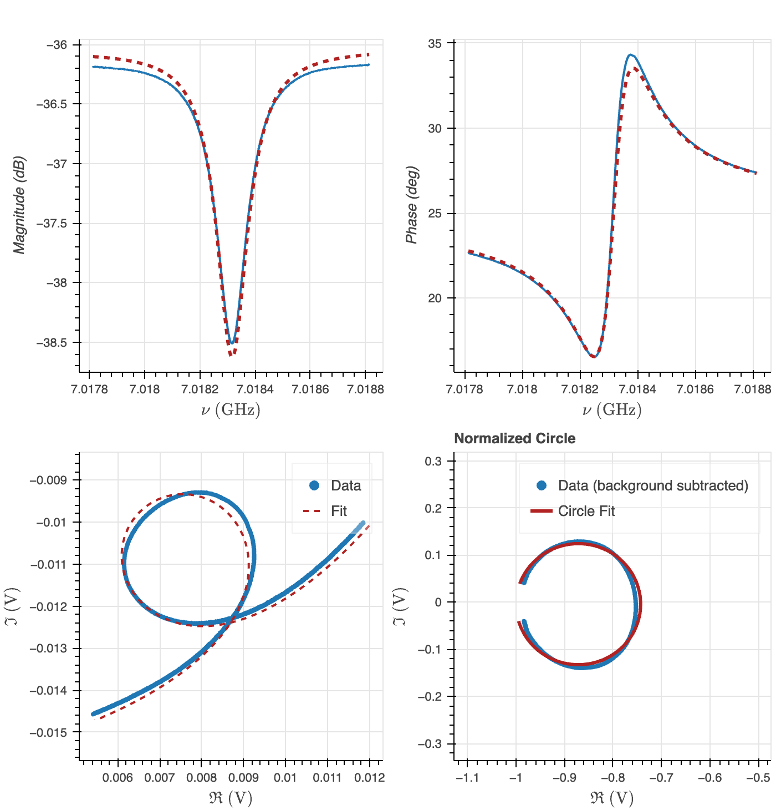}
        \label{fig:CircleFitNoDecompNoEllipt}
    \end{subfigure}
    \quad
    \begin{subfigure}[t]{0.48\linewidth}
        \centering
        \subcaption{\raisebox{-20ex}[0pt][0pt]{\hspace{130pt}}}
        \includegraphics[width=1\linewidth]{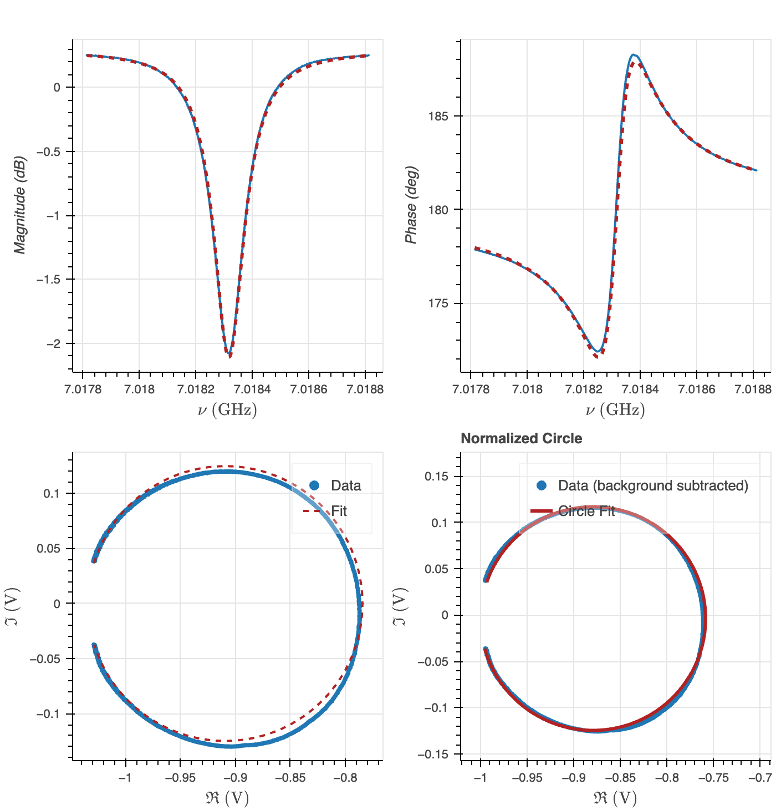}
        \label{fig:CircleFitDecompNoEllipt}
    \end{subfigure}
    \caption{Examples of circle fit of $S_{11}\left(\nu\right)$ traces before (a, c) and after (b, d) the decompression routine application. In the case of an elliptic compression and a coupling coefficient $\beta>0.5$, the correction method completely accounts for the amplifier compression, while for traces showing non-elliptic compression and a coupling coefficient $\beta<0.5$ the method is no longer effective.}    \label{fig:CircleFitExamplesWW/Decompression}
\end{figure}

\supplementarynote{Compensation for amplifier gain compression}
When sweeping the microwave power, the high-power data points fell in the compression region of the HEMT amplifier. As a result, the corresponding $S_{11}\left(\nu\right)$ appeared slightly distorted, resembling an ellipse more than a circle (Suppl.~Figs.~\ref{fig:CircleFitNoDecompEllipt}~-~\ref{fig:CircleFitNoDecompNoEllipt}). To correct the compression, the circle fit routine was slightly modified to include a regression to an ellipse. Starting from the shape of the background-subtracted reflection trace:

\begin{equation}
\tag{S2}
\label{eq:CompS11Nu}
S_{11}\left(\xi\right) = 1-\alpha\frac{1+2i Q_{l}\xi}{1+4Q_{l}^{2}\xi^{2}},
\end{equation}
with $\alpha=\frac{2Q_{l}}{Q_{ext}}, Q_{ext} \in \mathbb{C}$ and $\xi=\frac{\nu-\nu_{0}}{\nu_{0}}$, using the coordinate transformation to express the imaginary part as a function of the real part 

\begin{equation}
\tag{S3}
\label{eq:CoordMapCircle}
    \begin{cases}
        x\left(\xi\right) =1-\frac{\alpha}{1+4Q_{l}^{2}\xi^{2}} \\
        y\left(\xi\right) = \frac{2\alpha Q_{l}\xi}{1+4Q_{l}^{2}\xi^{2}}
    \end{cases},
\end{equation}
it is straightforward to arrive at the expression of a circle of center $c =\left(\frac{\alpha-1}{2},0\right)$ and radius $r=\frac{\alpha+1}{2}$. If the circle is deformed into an ellipse by the amplifier compression, by modifying the coordinate transformation in Eq.~(\ref{eq:CoordMapCircle}) with two additional parameters identifying the ellipse's semi-axes, the relation in Eq.~(\ref{eq:CompS11Nu}) can be rewritten to account for the trace modification:

\begin{equation}
\tag{S4}
\label{eq:CoordMapEllipse}
    \begin{cases}
        X\left(\xi\right) =\frac{x}{a} =\frac{1}{a}\left( 1-\frac{\alpha}{1+4Q_{l}^{2}\xi^{2}}\right) \\
        Y\left(\xi\right) = \frac{y}{b} = \frac{2\alpha Q_{l}\xi}{b\left(1+4Q_{l}^{2}\xi^{2}\right)},
    \end{cases}
\end{equation}
where $a,b$ are the ellipse's semi-axes. This allows us to include the elliptic feature into the curve parametric equation as a function of the normalized frequency $\xi$. The resulting modified reflection trace expression is 

\begin{equation}
\tag{S5}
\label{eq:CompS11NuEllipse}
\tilde{S}_{11}\left(\xi\right) = \frac{1}{a}-\alpha \left(\frac{b-2iaQ_{l}\xi}{ab\left(1+4Q_{l}^{2}\xi^2\right)}\right).
\end{equation}

The decompression routine is effective as long as the amplifier compression acts on the trace by deforming it into an ellipse, that is, with a precise relation between the real and imaginary parts or, equivalently, between the magnitude and phase of the trace (Suppl.~Figs.~\ref{fig:CircleFitNoDecompEllipt} and \ref{fig:CircleFitDecompEllipt}). However, if the compression shows a complex relation between signal magnitude and phase, the method cannot be  fully applied (Suppl.~Figs.~\ref{fig:CircleFitNoDecompNoEllipt} and \ref{fig:CircleFitDecompNoEllipt}). In addition, the correction routine is more effective with a critically coupled or slightly over/under-coupled antenna, whereas its performance significantly worsens outside of this regime, in accordance with the optimal operational regime for the original circle fit algorithm. 

\begin{figure}[h!] 
     \centering
     \includegraphics[width=1\textwidth]{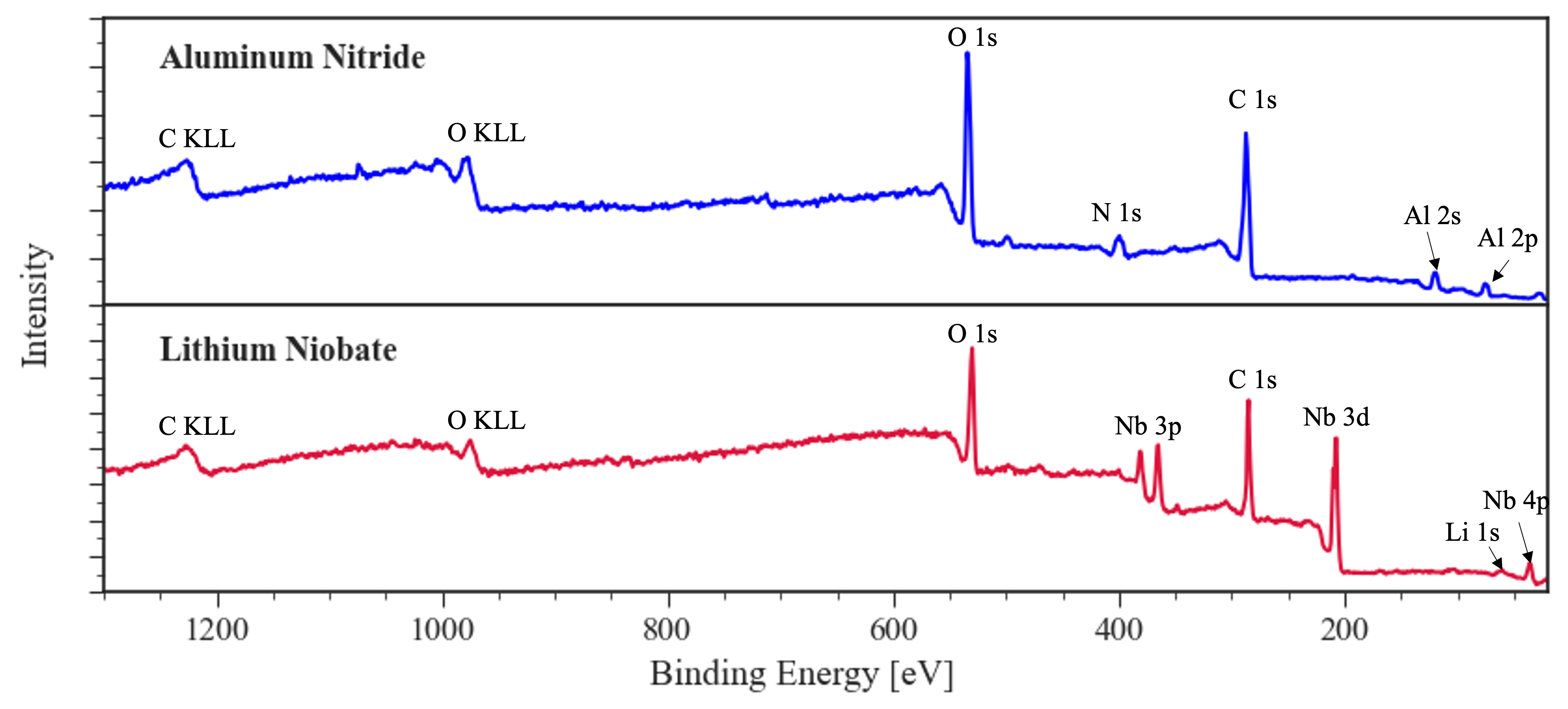}
    \caption{XPS survey spectra of AlN and LiNbO$_{3}$ crystals. Characteristic elemental peaks of each crystal are labeled, along with impurity signals from carbon and oxygen.}
    \label{fig: XPS Survey}
\end{figure}

\begin{figure}[htbp!]
  \centering
  \begin{subfigure}[t]{0.48 \linewidth}
    \centering
    \subcaption{\raisebox{0ex}[0pt][0pt]{\hspace{200pt}}}
    \includegraphics[width=1.05\linewidth]{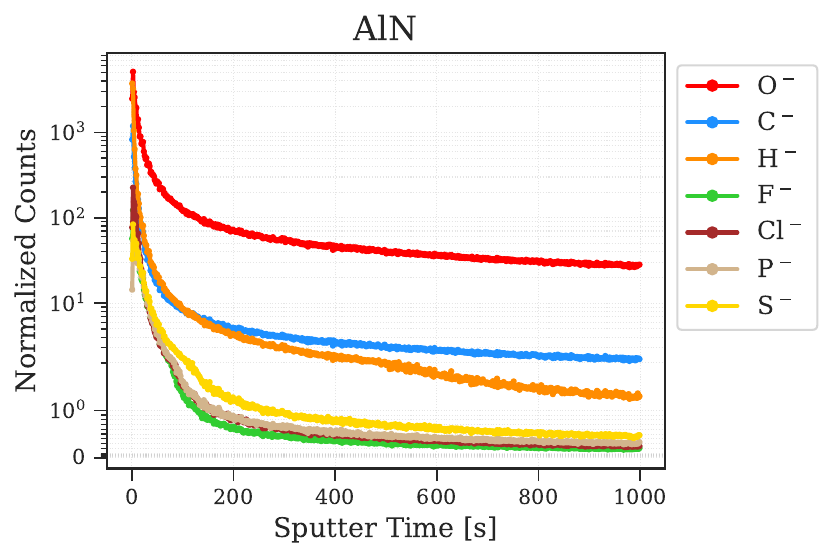}
    \label{fig:AlN Depth Profile}
  \end{subfigure}
  \quad
  \begin{subfigure}[t]{0.48 \linewidth}
    \centering
    \subcaption{\raisebox{0ex}[0pt][0pt]{\hspace{200pt}}}
    \includegraphics[width=1.05\linewidth]{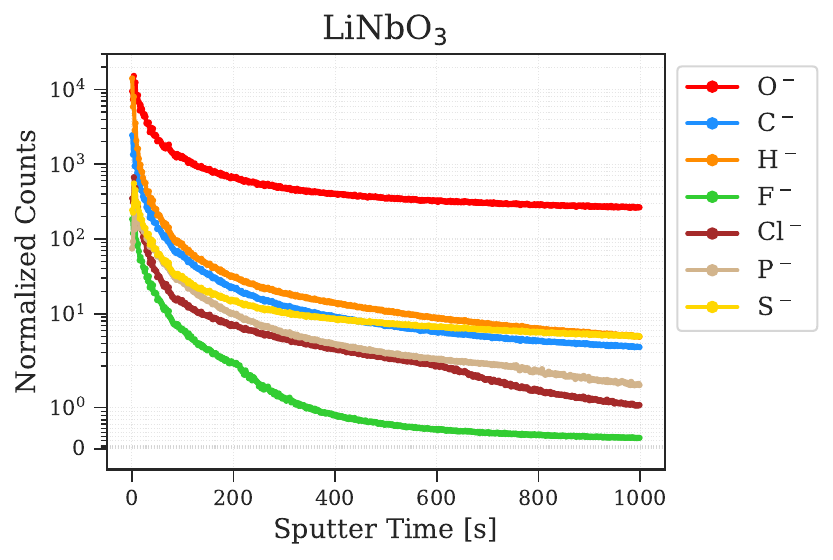}
    \label{fig:LiNbO3 Depth Prifle}
  \end{subfigure}
  \caption{TOF-SIMS depth profiles of normalized impurity signals for (a) AlN and (b) LiNbO$_3$. The plots show the evolution of impurity counts as a function of sputter time, which is directly related to the depth into the crystal.}
    \label{fig:SIMS Impurities depth profile}
\end{figure}

\supplementarynote{Material analysis at room temperature}

Each of the two bulk crystals was investigated using XPS and TOF-SIMS. Supplementary Figure~\ref{fig: XPS Survey} shows the XPS survey spectra of the bulk crystals, which reveal the elemental composition within the top ~5–10 nm of the surface. All binding energies were referenced to the C 1s hydrocarbon peak at 285 eV. Both spectra display prominent peaks for oxygen and carbon, along with their corresponding Auger counterparts. The carbon peak originates from adventitious surface contamination, while the intense oxygen signal observed in the AlN crystal indicates the formation of an aluminum oxide layer at the surface. In both the AlN and LiNbO$_3$ spectra, the intrinsic lattice peaks (e.g., Al 2p and N 1s for AlN; Li 1s and Nb 3d for LiNbO$_3$) appear with relatively lower intensity, consistent with partial surface contamination. In general, the detection limits of XPS are on the order of ~1 at.\% down to ~0.1 at.\% for most elements \cite{shard2014detection}. As such, the technique is not well-suited for identifying trace impurities below this threshold.

\begin{figure}[htbp!]
    \centering
    \begin{subfigure}{0.7\textwidth}
        \centering
        \subcaption{\raisebox{0ex}[0pt][0pt]{\hspace{425pt}}}
        \includegraphics[width=0.9\linewidth]{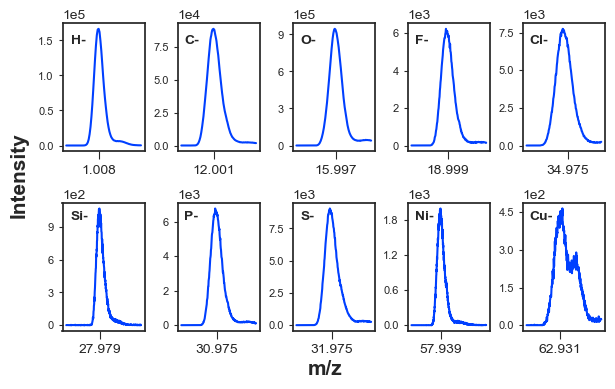}
        \label{fig:Q0vsTLiNbO3TM010}
    \end{subfigure}
    \\
    \begin{subfigure}{0.7\textwidth}
        \centering
        \subcaption{\raisebox{0ex}[0pt][0pt]{\hspace{425pt}}}
        \includegraphics[width=0.9\linewidth]{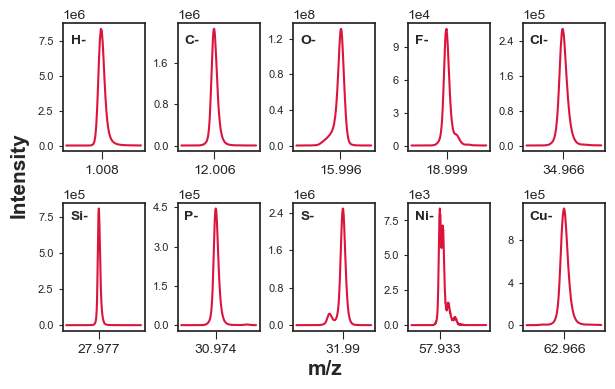}
        \label{fig:Q0vsTLiNbO3TE011}
    \end{subfigure}
    \caption{TOF-SIMS mass spectra (intensity vs m/z) of various contaminants found on the surface and within the bulk of (a) AlN (blue line plots) and (b) LiNbO$_3$ (red line plots) crystals. }
    \label{fig:SIMS Impurity Signals}
\end{figure}

To further investigate surface and bulk impurities, TOF-SIMS analysis was carried out on both crystals. Supplementary Figures~\ref{fig:AlN Depth Profile} and ~\ref{fig:LiNbO3 Depth Prifle} show the depth profiles, where the y-axis shows the impurity counts normalized to Al$^-$ in AlN and Nb$^-$ in LiNbO$_3$, respectively. The impurity signals decrease with sputter time, reflecting their higher concentration at the surface and reduced levels deeper within the bulk. This indicates that the contaminant species are largely confined to the surface region and are negligible in the bulk of either crystal. Supplementary Figure~\ref{fig:SIMS Impurity Signals} displays the negative ion mass spectra (intensity vs. m/z) for selected impurity species, including O$^-$, C$^-$, H$^-$, F$^-$, Cl$^-$, Si$^-$, P$^-$, S$^-$, Ni$^-$, and Cu$^-$, obtained from (a) AlN and (b) LiNbO$_3$. In both crystals, signals indicative of H$^-$, C$^-$, and O$^-$ dominate the spectra, consistent with surface contamination from hydrocarbons, adsorbed water, and air exposure. Additional impurities, such as F$^-$ and Cl$^-$, are also detected and likely originate from polishing, etching, or handling. Weaker peaks corresponding to Si$^-$, P$^-$, and S$^-$ suggest the presence of trace environmental contaminants \cite{soffey2022analysis}. The Ni$^-$ and Cu$^-$ weak signals indicate metallic impurities introduced through tools or environmental contact.

\supplementarynote{Extended power and temperature sweep results with relative frequency shift analysis}
The purpose of this section is to provide a thorough report of the regression results for both power sweep and temperature sweep, as well as to perform a qualitative analysis of the temperature-induced resonance frequency shift. Base-temperature power regression results following the models in the main text for the two resonances of interest are reported in Suppl. Tabs.~\ref{tab:PwrSweepRegrResultsTM010} and \ref{tab:PwrSweepRegrResultsTE011}.

\begin{figure}[htbp!]
    \centering
    \begin{subfigure}{0.45\textwidth}
        \centering
        \subcaption{\raisebox{0ex}[0pt][0pt]{\hspace{425pt}}}
        \includegraphics[width=0.95\linewidth]{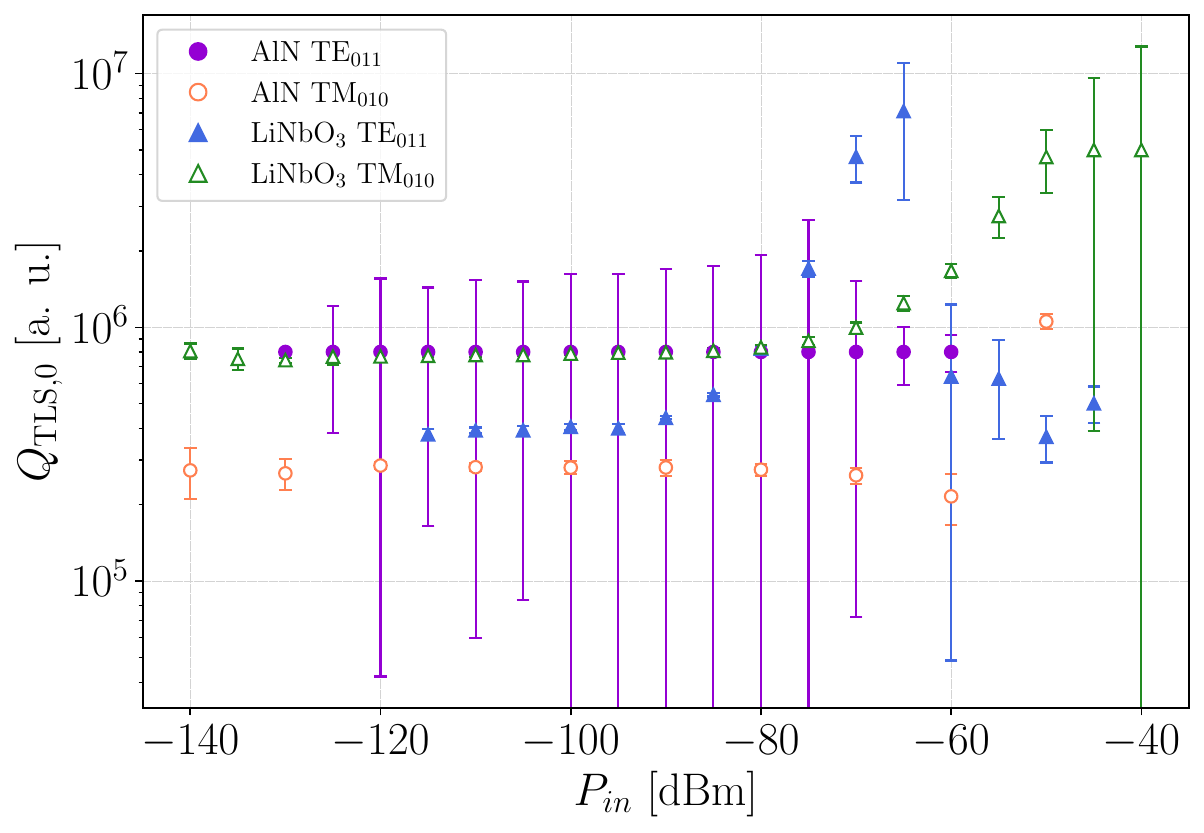}
        \label{fig:QTLS0Aggr}
    \end{subfigure}
    \quad
    \begin{subfigure}{0.45\textwidth}
        \centering
        \subcaption{\raisebox{0ex}[0pt][0pt]{\hspace{425pt}}}
        \includegraphics[width=0.95\linewidth]{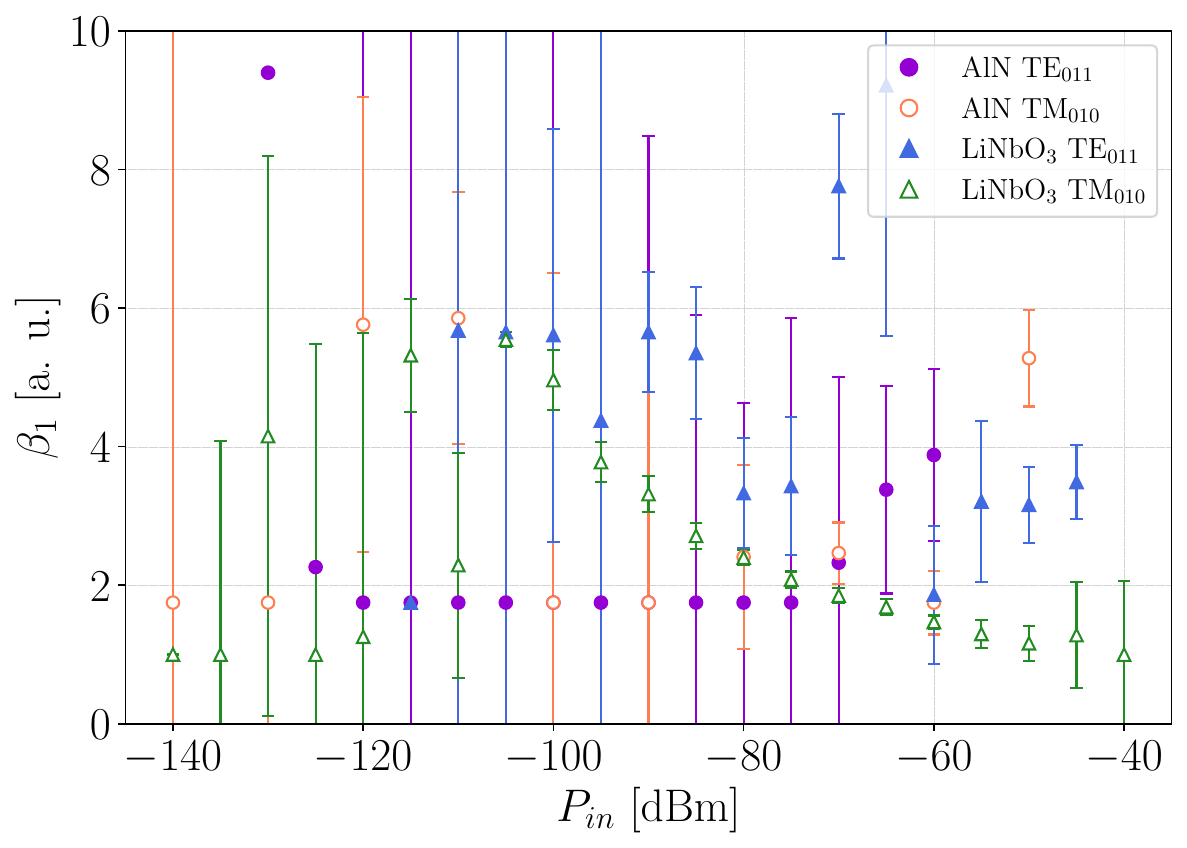}
        \label{fig:Beta1Aggr}
    \end{subfigure}
    \\
    \begin{subfigure}{0.45\textwidth}
        \centering
        \subcaption{\raisebox{0ex}[0pt][0pt]{\hspace{425pt}}}
        \includegraphics[width=0.95\linewidth]{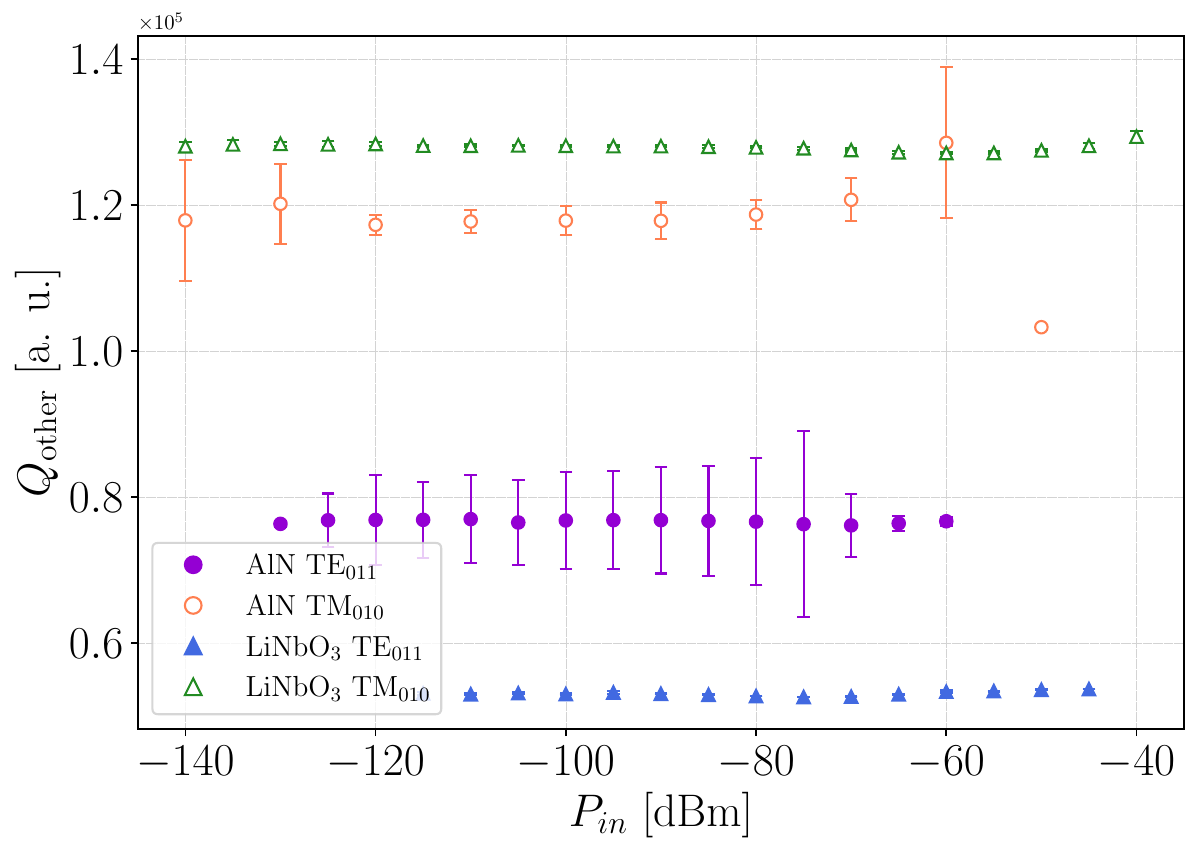}
        \label{fig:QotherAggr}
    \end{subfigure}
    \quad
    \begin{subfigure}{0.45\textwidth}
        \centering
        \subcaption{\raisebox{0ex}[0pt][0pt]{\hspace{425pt}}}
        \includegraphics[width=0.95\linewidth]{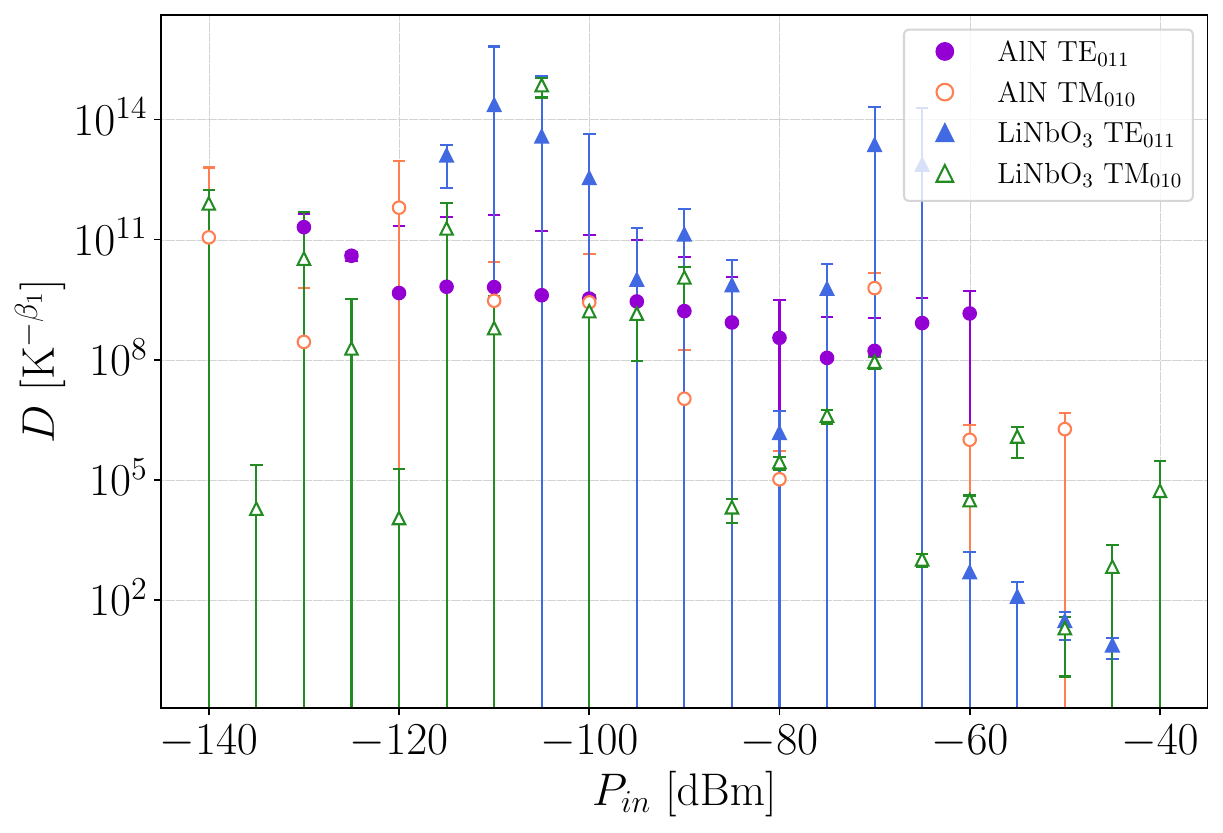}
        \label{fig:DAggr}
    \end{subfigure}
    \\
    \begin{subfigure}{0.45\textwidth}
        \centering
        \subcaption{\raisebox{0ex}[0pt][0pt]{\hspace{425pt}}}
        \includegraphics[width=0.95\linewidth]{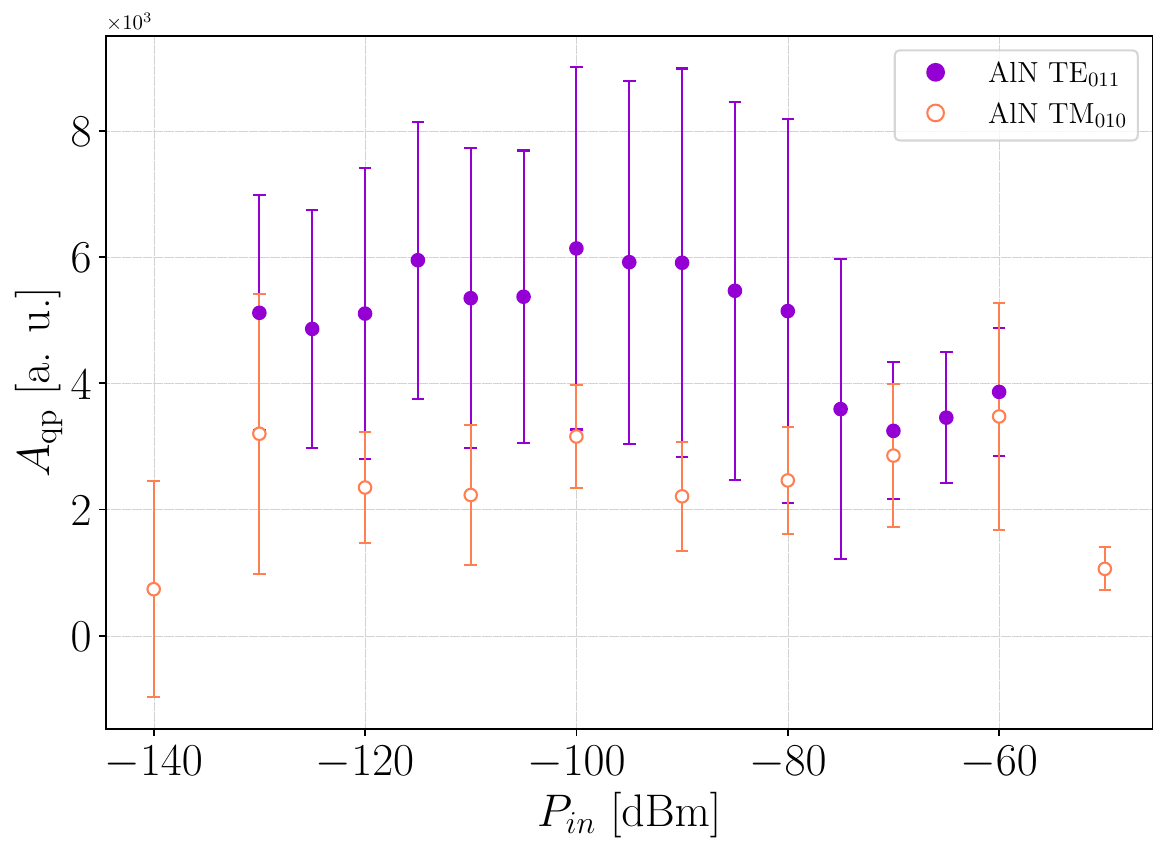}
        \label{fig:AqpAggr}
    \end{subfigure}
    \quad
    \begin{subfigure}{0.45\textwidth}
        \centering
        \subcaption{\raisebox{0ex}[0pt][0pt]{\hspace{425pt}}}
        \includegraphics[width=0.95\linewidth]{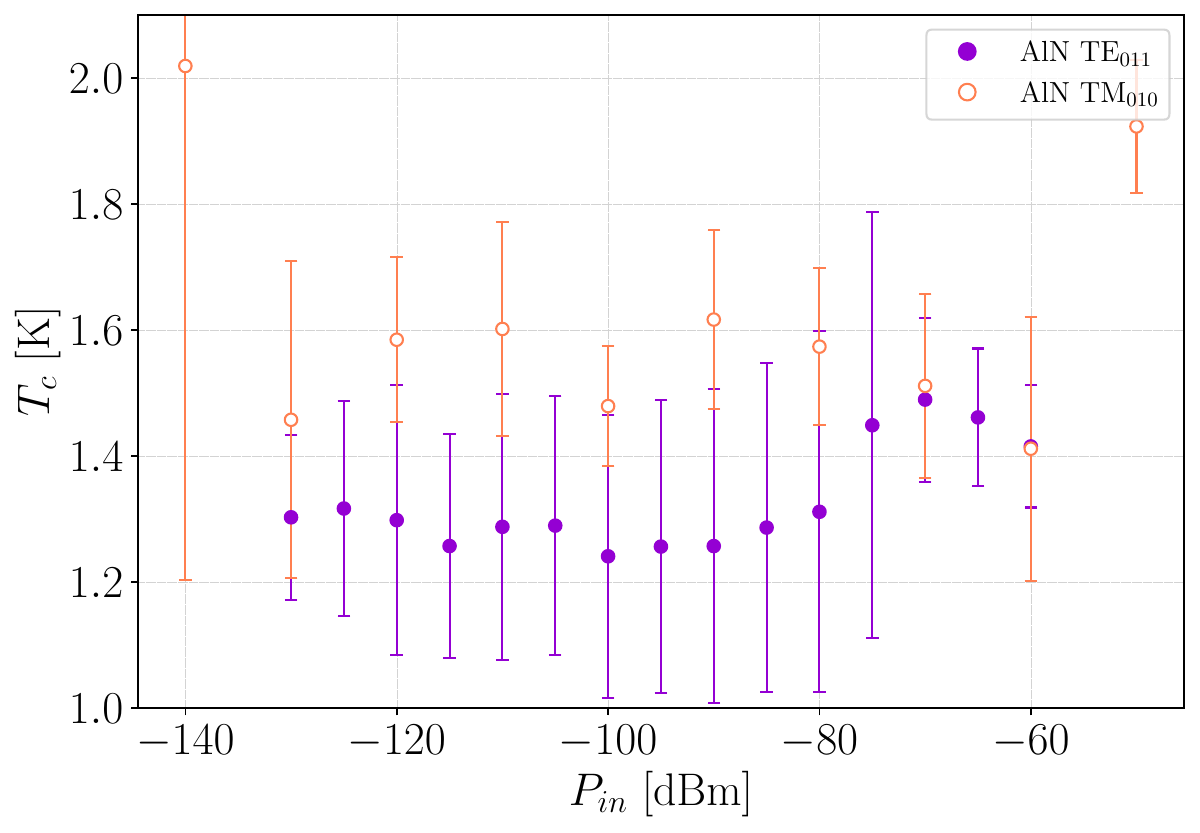}
        \label{fig:TcritAggr}
    \end{subfigure}
    \caption{Aggregated plots of the temperature sweep regression parameters for each mode of both materials versus input power at the cavity port. As mentioned in the main text, the TLS quality factor at base temperature $Q_\mathrm{TLS,0}$ (a) is constant for power magnitudes smaller than -60 dBm for all resonances under study. (b) Exponent for inverse temperature quality factor dependence $\beta_{1}$, varying considerably with power for the TE$_{011}$ modes, less for the TM$_{010}$ modes. (c) Quality factor from other sources of loss  $Q_\mathrm{other}$, almost constant at any power level for all materials and compatible with the values extracted from power sweep regressions. (d) Amplitude of the inverse temperature quality factor dependence $D$. (e) Kinetic inductance due to quasiparticles in the cavity walls $A_\mathrm{qp}$ and superconducting critical temperature $T_{c}$ for the AlN resonances.}
    \label{fig:TsweepParamAggr}
\end{figure}

\input{Tables/Power_sweep_regression_results_TM010_TE011_full}
For the temperature measurements, the parameters extracted from regressions are plotted as a function of the input power at the entrance of the cavity  (Suppl. Fig.~\ref{fig:TsweepParamAggr}).

Finally, we analyze the materials' resonance frequency shift as a function of the temperature normalized by the resonance frequency at base temperature $\nu_{0}$. The model comprises two parts to describe the frequency shift~\cite{crowley2023disentangling,gao2008thephysics}:

\begin{equation}
    \tag{S6}
    \label{eq:DeltaNuvsTglob}
    \frac{\delta\nu\left(T\right)}{\nu_{0}} = \left(\frac{\delta\nu\left(T\right)}{\nu_{0}}\right)_{\mathrm{TLS}} + \left(\frac{\delta\nu\left(T\right)}{\nu_{0}}\right)_{\mathrm{QP}},
\end{equation}
The first is a TLS-dependent component:

\begin{equation}
   \tag{S7}
   \label{eq:DeltaNuvsTTLS}
   \left(\frac{\delta\nu\left(T\right)}{\nu_{0}}\right)_{\mathrm{TLS}} = \frac{1}{\pi Q_{\mathrm{TLS,0}}}\mathfrak{R}\left[\Psi\left(\frac{1}{2} + i\frac{h\nu_{0}}{2\pi k_{B}T}\right)-\log\left(\frac{h\nu_{0}}{2\pi k_{B}T}\right)\right],
\end{equation}
where $\Psi$ is the Euler digamma function and $Q_{\mathrm{TLS,0}}$ is, once again, the zero-temperature TLS quality factor. The second term is a quasiparticle component that takes into account Cooper pair breaking at higher temperatures, resulting in conducting electrons in the walls of the cavity:

\begin{equation}
    \tag{S8}
    \left(\frac{\delta\nu\left(T\right)}{\nu_{0}}\right)_{\mathrm{QP}} = -\frac{\alpha}{2}\left(\frac{\pi\Delta_{0}}{h\nu_{0}}\frac{1}{\left\|\sigma\left(T,\nu_{0}\right)\right\|}\sin\left(\phi\left(T,\nu_{0}\right)\right)-1\right),
\end{equation}
where $\sigma = \sigma_{1} +i\sigma_{2}$ is the classical electrons complex conductivity $\phi\left(T,\nu_{0}\right)=\tan\left(\frac{\sigma_{2}\left(T,\nu_{0}\right)}{\sigma_{1}\left(T,\nu_{0}\right)}\right)$, $\Delta_{0}$ is, once again, the superconducting bandgap and $\alpha$ the fraction of kinetic inductance over total inductance of the cavity walls \cite{gao2008thephysics,gao2008equivalence}. The fitting parameters for this model are $Q_{\mathrm{TLS,0}}$, $\alpha$ and $T_{c}$.

\begin{figure}[htbp!]
    \centering
    \begin{subfigure}{0.9\textwidth}
        \centering
        \subcaption{\raisebox{0ex}[0pt][0pt]{\hspace{425pt}}}
        \includegraphics[width=0.95\linewidth]{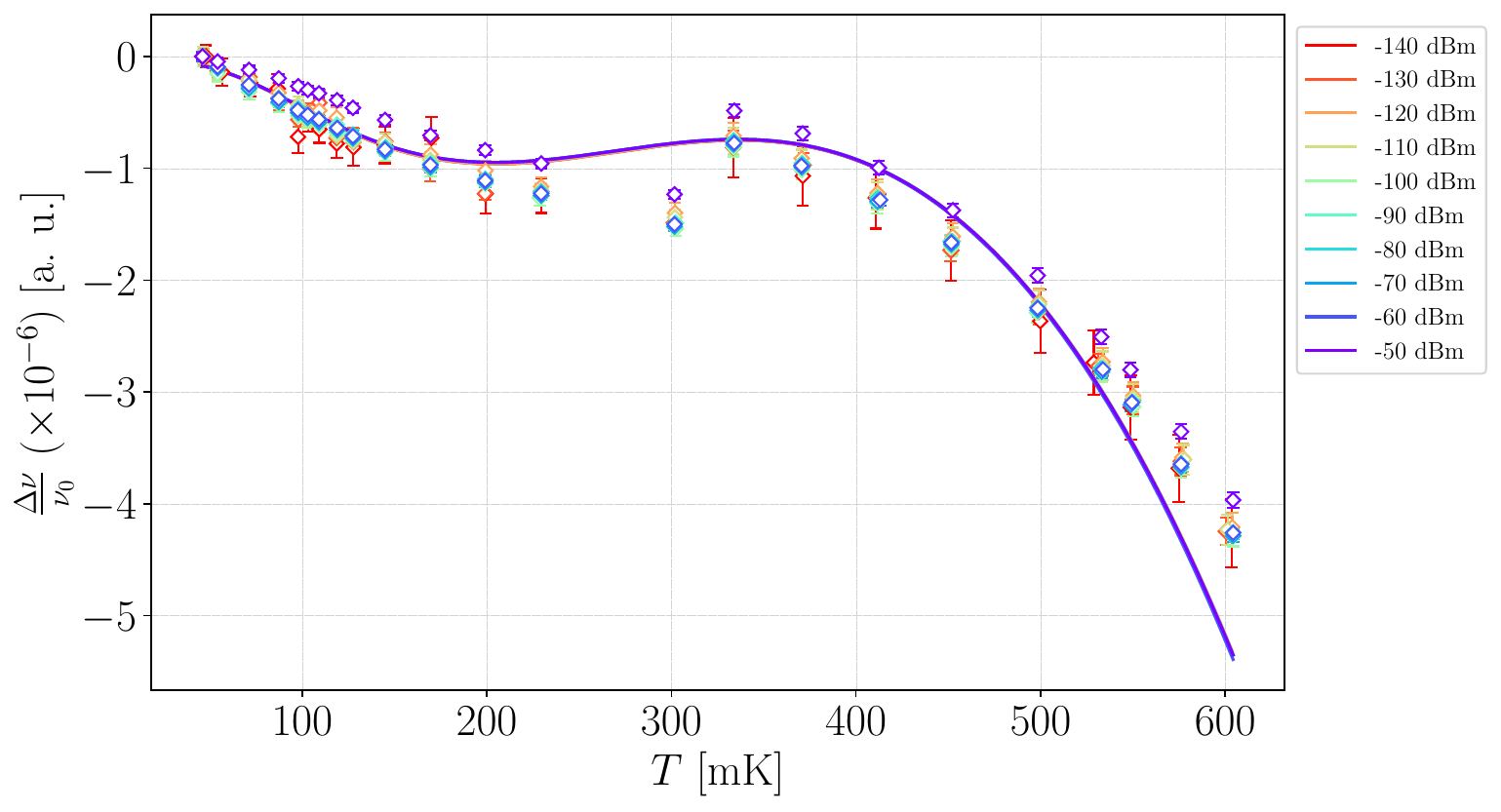}
        \label{fig:DeltaNuRelVsTAlNTM010}
    \end{subfigure}
    \\
    \begin{subfigure}{0.9\textwidth}
        \centering
        \subcaption{\raisebox{0ex}[0pt][0pt]{\hspace{425pt}}}
        \includegraphics[width=0.95\linewidth]{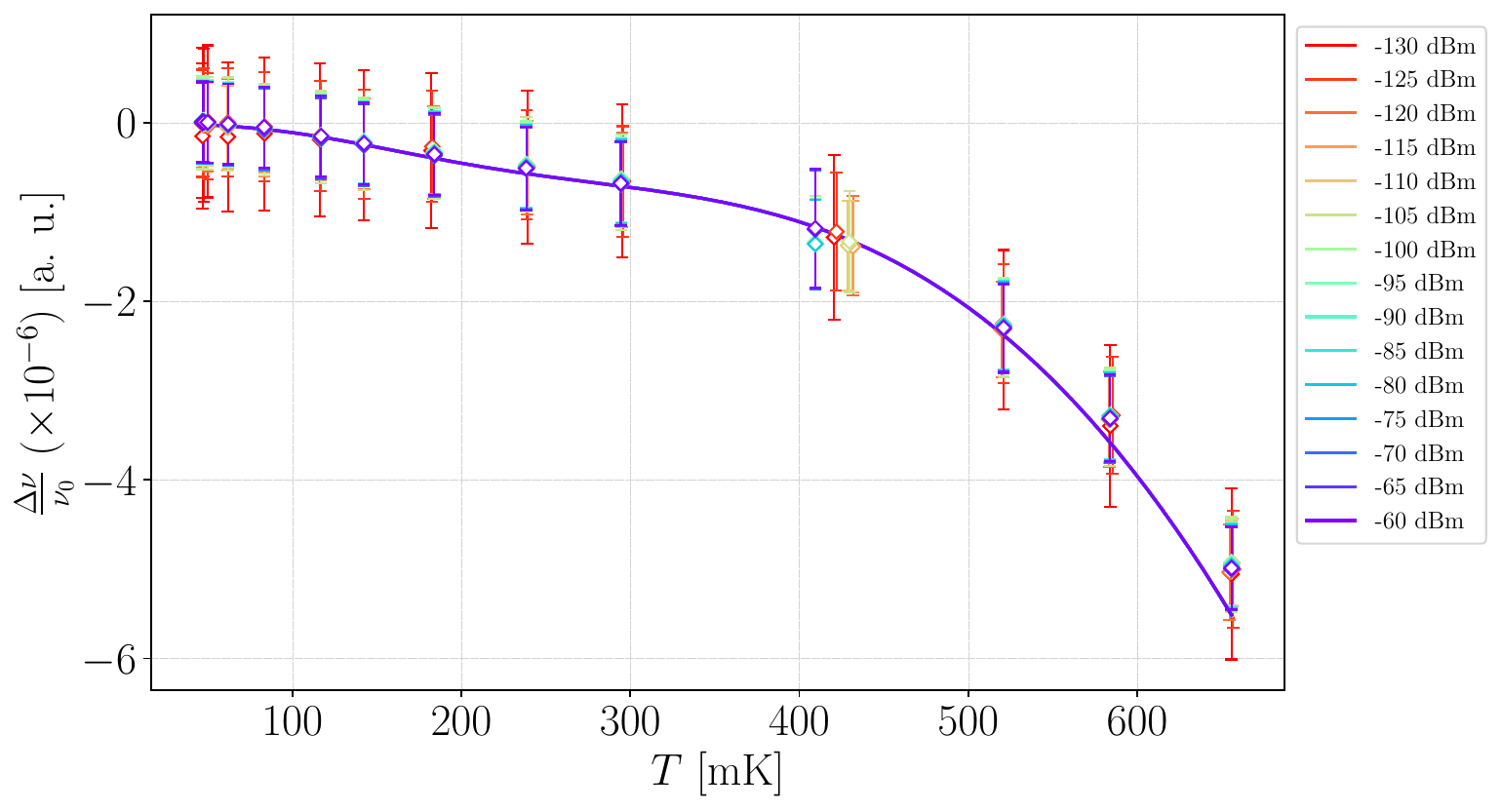}
        \label{fig:DeltaNuRelVsTAlNTE011}
    \end{subfigure}
    \caption{Relative frequency shift versus temperature for TM$_{010}$ (a) and TE$_{011}$ (b) modes of AlN with corresponding regressions. The fitted curves show a TLS-dominated behavior for $T<400$ mK and a change in convexity with the onset of quasiparticles at higher temperatures.}
    \label{fig:DeltaNuRelVsTAlN}
\end{figure}

The analysis of the relative frequency shift within the same temperature range provides further evidence of the TLS nature of the dielectric losses in the material in the case of the AlN sample resonances. The model proposed in Eq.~(\ref{eq:DeltaNuvsTglob}) fits the data with good agreement for temperatures below 500~mK, clearly showing a TLS-dominated behavior for lower T values ($T<400$ mK), replaced by a quasiparticle-related behavior as temperature increases (Fig.~\ref{fig:DeltaNuRelVsTAlN}). The extracted values of superconducting critical temperature $T_{c}$ from the regressions, namely 1.2~K for the transverse magnetic mode and 1.3~K for the transverse electric mode, are compatible with the critical temperature of bulk aluminum. However, the retrieved base-temperature TLS quality factor is underestimated by the regression by one order of magnitude.

\begin{figure}[htbp!]
\centering
    \begin{subfigure}{1\textwidth}
        \centering
        \subcaption{\raisebox{0ex}[0pt][0pt]{\hspace{425pt}}}
        \includegraphics[width=0.9\linewidth]{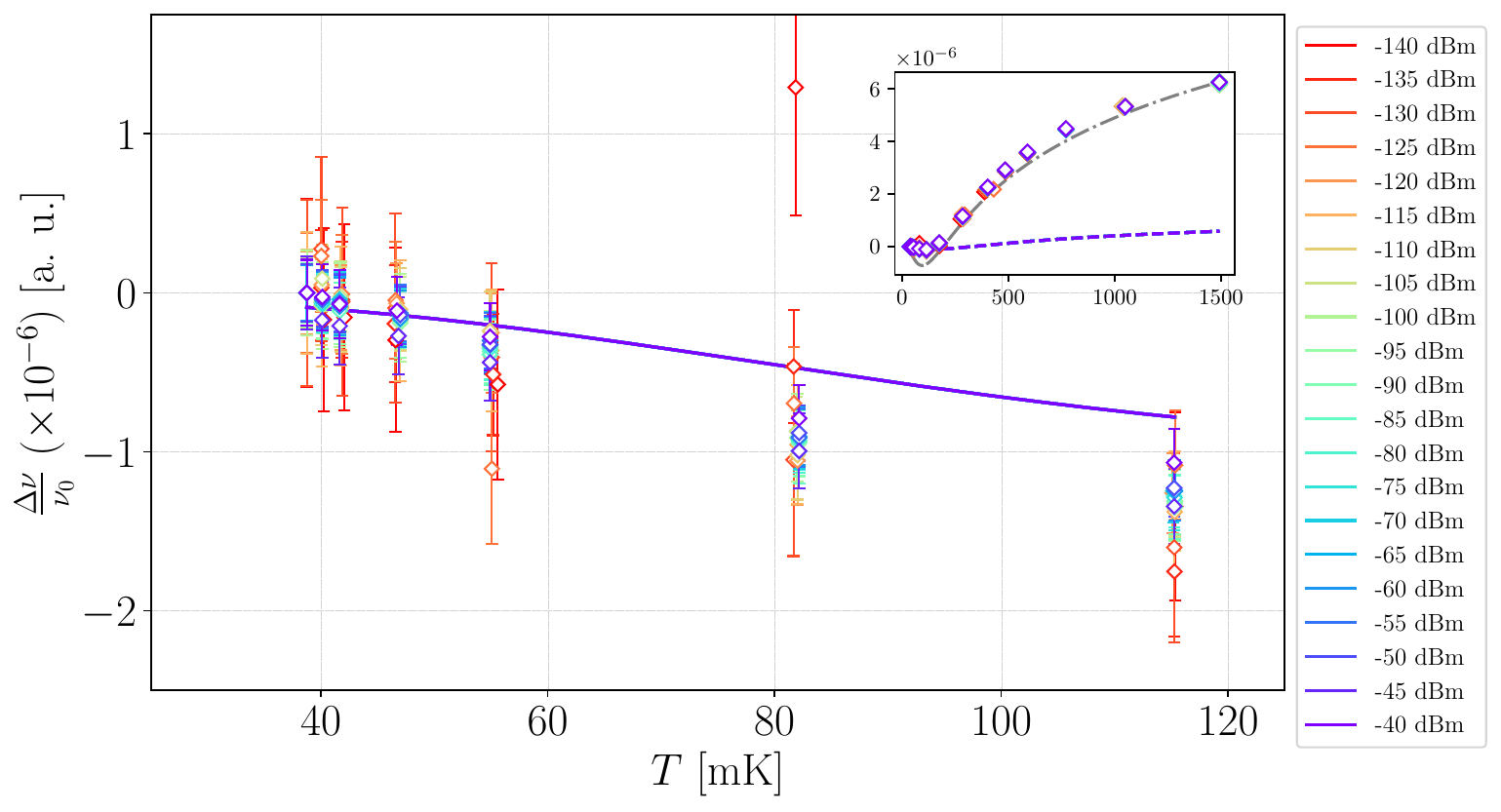}
        \label{fig:DeltaNuRelVsTLiNbO3TM010}
    \end{subfigure}
    \\
    \begin{subfigure}{0.9\textwidth}
        \centering
        \subcaption{\raisebox{0ex}[0pt][0pt]{\hspace{425pt}}}
        \includegraphics[width=0.95\linewidth]{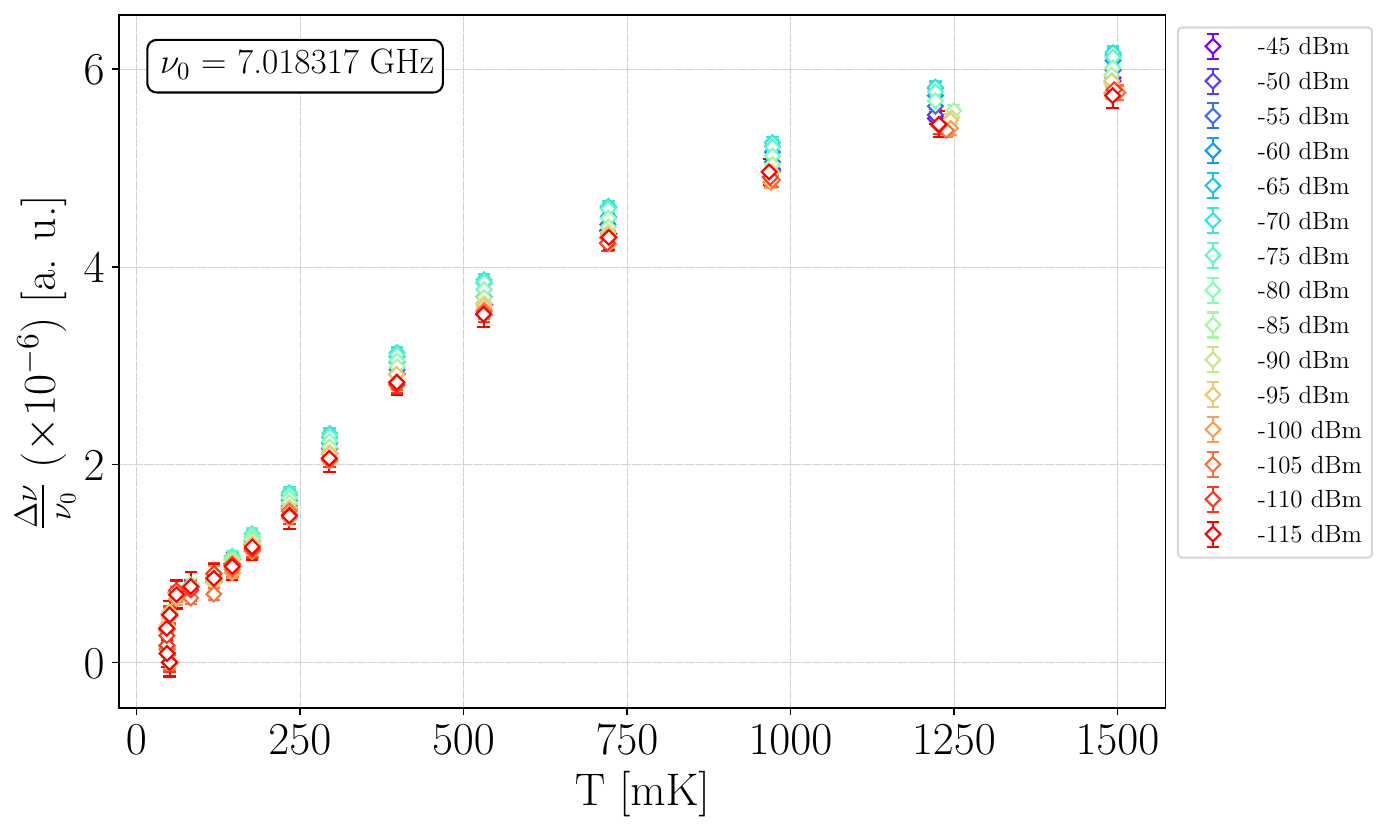}
        \label{fig:DeltaNuRelVsTLiNbO3TE011}
    \end{subfigure}
    \caption{Relative frequency shift versus temperature for TM$_{010}$ (a) and TE$_{011}$ (b) modes of LiNbO$_{3}$. Regressions are performed on the transverse magnetic resonance data set only, with the model in Eq.~(\ref{eq:DeltaNuvsTTLS}) since the crystal is inserted in a Nb cavity. The relative frequency shift data follow the model for the first 100 mK of temperature span, then for values of $Q_{\mathrm{TLS,0}}$ compatible with the one extracted from base temperature power sweep, they diverge significantly for higher temperature values (inset of b, colored solid line). Agreement between data series and regression is recovered for $Q_{\mathrm{TLS,0}}\sim10^{3}$ (inset of (a), dot-dashed gray line). The frequency shift of the transverse electric resonance (b) is monotonically increasing at any explored temperature and not compatible with either frequency shift model.}
    \label{fig:DeltaNuRelVsTLiNbO3}
\end{figure}

The LiNbO$_{3}$ sample relative frequency shift showed a more peculiar behavior overall. For the TM$_{010}$ mode, the model in Eq.~(\ref{eq:DeltaNuvsTTLS}) would only describe the entire data series well if $Q_\mathrm{TLS,0}$ were decreased to $10^{3}$, while with a quality factor compatible with temperature sweep regression results the curve would only fit the dataset from 10 mK to 150 mK (Fig.~\ref{fig:DeltaNuRelVsTLiNbO3TM010}). As for the transverse electric resonance, the shift increased monotonically over the whole temperature range, which is incompatible with the regression model (Fig.~\ref{fig:DeltaNuRelVsTLiNbO3TE011}).

\bibliographystyle{naturemag}
\bibliography{references}

%% file: abstract.tex
Efficient bidirectional microwave-optical photon conversion is a key capability for scaling superconducting quantum processors into distributed networks. However, achieving the necessary conversion efficiency requires filling a critical knowledge gap in understanding the loss mechanisms of electro-optic materials. Here, we characterize the microwave properties of single-crystal bulk LiNbO$_{3}$ and AlN over a broad range of powers, down to single-photon levels, and spanning from millikelvin temperatures to above 1~K. We demonstrate that both materials exhibit two-level systems (TLS) behavior, while piezoelectric-related losses are excluded. We show that TLS-induced dissipation is predominantly localized on the surface rather than being an intrinsic bulk property, a result further corroborated by room-temperature 3D XPS and time-of-flight SIMS analyses. These findings provide useful insights to engineer hybrid architectures that integrate bulk electro-optic crystals within superconducting cavities, proving that microwave quality factors compatible with high-efficiency microwave–optical transduction are within reach.

%% file: introduction.tex
\label{sec:introduction}
The past decade has witnessed remarkable advancements in quantum computing based on superconducting platforms~\cite{schoelkopf2008wiring, blais2020circuit, wendin2017quantum}.
As superconducting quantum processing units continue to scale up, so does the demand for efficient cooling solutions capable of sustaining cryogenic operating temperatures for larger devices. This trend highlights the need for hybrid solutions, interconnecting processing units and quantum networks. Infrared (IR) photons are resistant to thermal noise and have proven to be a promising candidate for long-distance quantum communication~\cite{pirandola2016physics, zhou2023experimental}. IR photons operate at frequencies of hundreds of terahertz (THz), while superconducting quantum devices operate in the gigahertz (GHz) regime. Microwave-optical quantum transducers bridge this energy gap with low added noise~\cite{lauk2020perspectives, han2021microwave, wang2022high-efficiency, bargagna2025design, wu2025entangling}. Among the available implementations of microwave-optical transduction, frequency conversion via electro-optic materials offers a promising solution. In such materials, information conversion is mediated by modulation of the refractive index through an electromagnetic field. This results in a three-wave mixing process that converts photons at microwave frequencies into IR with the help of a laser pump source. The interaction of microwave and optical modes is based on the second-order nonlinearity of the electric susceptibility of such materials, which is described by the tensor $\chi^{\left(2\right)}_{mnl}$~\cite{wang2022high-efficiency, tsang2010cavity, tsang2011cavity}. 

Electro-optic materials such as lithium niobate (LiNbO$_{3}$) and aluminum nitride (AlN) have been used to demonstrate quantum microwave-optical bidirectional conversion~\cite{rueda2016efficient,xu2021bidirectional}. Optical cavities made of these electro-optic materials exhibit a high quality factor $Q_\mathrm{opt}$ for the optical mode. In hybrid systems, instead, the quality factor of the microwave mode $Q_\mathrm{mw}$ remains significantly low, leading to limitations in the overall conversion efficiency~\cite{zhang2017monolithic, hease2020bidirectional, sahu2022quantum-enabled}. The limited $Q$ of the microwave mode is related to the high loss tangent of the crystal, i.e., the imaginary part of the permittivity, and is further limited by spurious loss mechanisms, often attributed to specific additional properties of the electro-optic materials. Such crystals exhibit piezoelectric effects~\cite{cho1987nonlinear,gui1999extensional,weis1985lithium}, ferroelectric effects~\cite{abrahams1966ferroelectric}, pyroelectric effects~\cite{sergeeva2021dielectric,shaldin2008pyroelectric} and anisotropic relative electric permittivity~\cite{yong1993electronic,ohmachi1967dielectric}. Detailed investigations of microwave losses induced by the mechanisms mentioned above have been carried out for thin films~\cite{yang2023piezoelectric,scigliuzzo2020phononic} and micrometer-scale resonators~\cite{wollack2021loss}. These studies have identified the presence of non-negligible piezoelectric-induced losses. While hybrid lithographic thin-film devices suffer from these reductions in $Q_\mathrm{mw}$, monolithic bulk crystal-cavity architectures provide evidence that substantially higher microwave quality factors may be achieved when the electro-optic crystals are embedded in superconducting radiofrequency (SRF) cavities, making them promising for high-efficiency quantum transduction~\cite{wang2022high-efficiency}. The extent of these spurious effects at the quantum level in macroscopic versions of the same crystalline materials is not yet well understood~\cite{zorzetti2023millikelvin, goryachev2015single-photon}. Exploring the complex microwave quantum behavior of electro-optic materials can reveal these interactions and inform strategies to mitigate undesirable coupling mechanisms, ultimately selectively enhancing the electro-optic effect and enabling higher-efficiency microwave–optical transducers with improved fidelity.

In this work, we investigate the microwave response of macroscopic single crystals of LiNbO$_{3}$ and AlN at cryogenic temperatures. The primary objective is to assess the extent to which the piezoelectric- and ferroelectric-associated microwave loss mechanisms, previously observed in thin-film and micrometer-scale structures, remain operative in macroscopic single crystals at millikelvin temperatures. Here, we directly measure the dependence of the internal microwave quality factor ($Q_{0}$) on power down to the single-photon level, at temperatures in the range from a few mK to above 1~K, and with different field orientations. To perform this study, we designed a custom three-dimensional SRF cavity, which provides a well-defined electromagnetic environment for the samples, allowing selective evaluation of excitation along different crystallographic directions of the material's electric permittivity depending on the mode analyzed. Both materials exhibit similar behavior, showing a power dependence of the quality factor that indicates the presence of two-level systems (TLS) as a loss mechanism at low microwave power. The TLS response is also confirmed by the temperature dependence of the internal quality factor $Q_{0}$, following the models in the literature~\cite{muller2019towards,crowley2023disentangling}. On the other hand, the measurements show no evidence of the presence of spurious piezoelectric coupling. Interestingly, saturation of TLS occurs at a relatively low average photon number in the cavity, suggesting that TLS with microwave energy splitting are only present on the outer surface of the samples. This finding is confirmed by 3D X-ray photoelectron spectroscopy (XPS) and time-of-flight secondary ion mass spectrometry (ToF SIMS) measurements at room temperature, performed to analyze the surface and bulk chemical and physical composition of the crystals. Finally, we highlight a peculiar difference between the $Q_{0}$ base temperature power measurements as a function of power in both the perpendicular and parallel directions. We observe that in both crystals  $Q_{0, \parallel}$ grows logarithmically, while $Q_{0, \perp}$ saturates at a constant value. This study opens new implementation scenarios centered around hybrid SRF cavity-electro-optic crystal packages with very promising quantum transduction efficiency.

%% file: results.tex
\subsection*{Measurement setup and technique}
\label{sec:MeasTechn}
\begin{figure}[htbp!]
    \begin{subfigure}[t]{0.52\linewidth}
        \centering
        \subcaption{\raisebox{0ex}[0pt][0pt]{\hspace{190pt}}}
        \includegraphics[scale = 0.3]{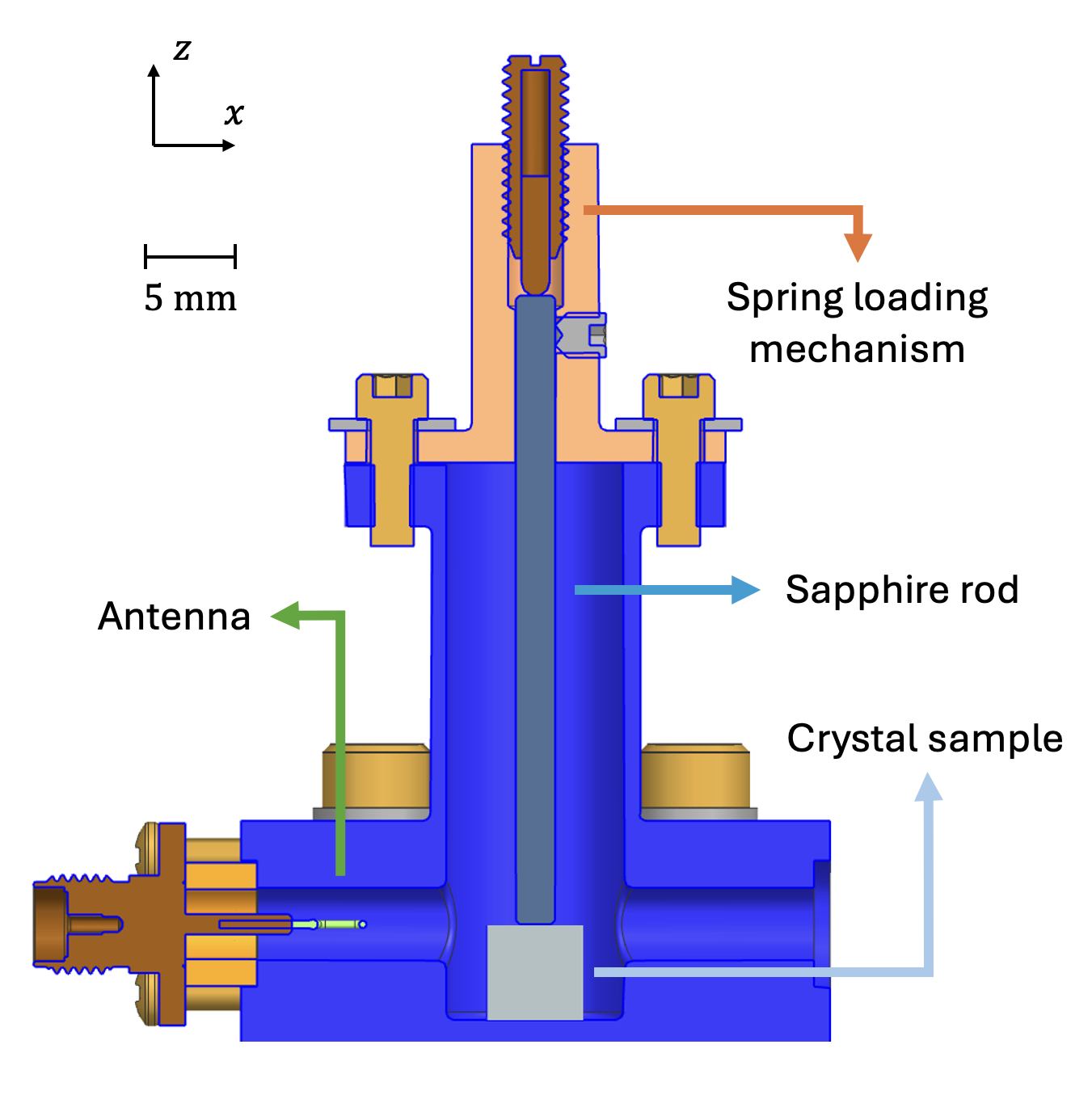}
        \label{fig:CavityCADSection}
    \end{subfigure}
    \quad
    \begin{subfigure}[t]{0.3\linewidth}
        \centering
        \subcaption{\raisebox{0ex}[0pt][0pt]{\hspace{20pt}}}
        \includegraphics[scale = 0.3]{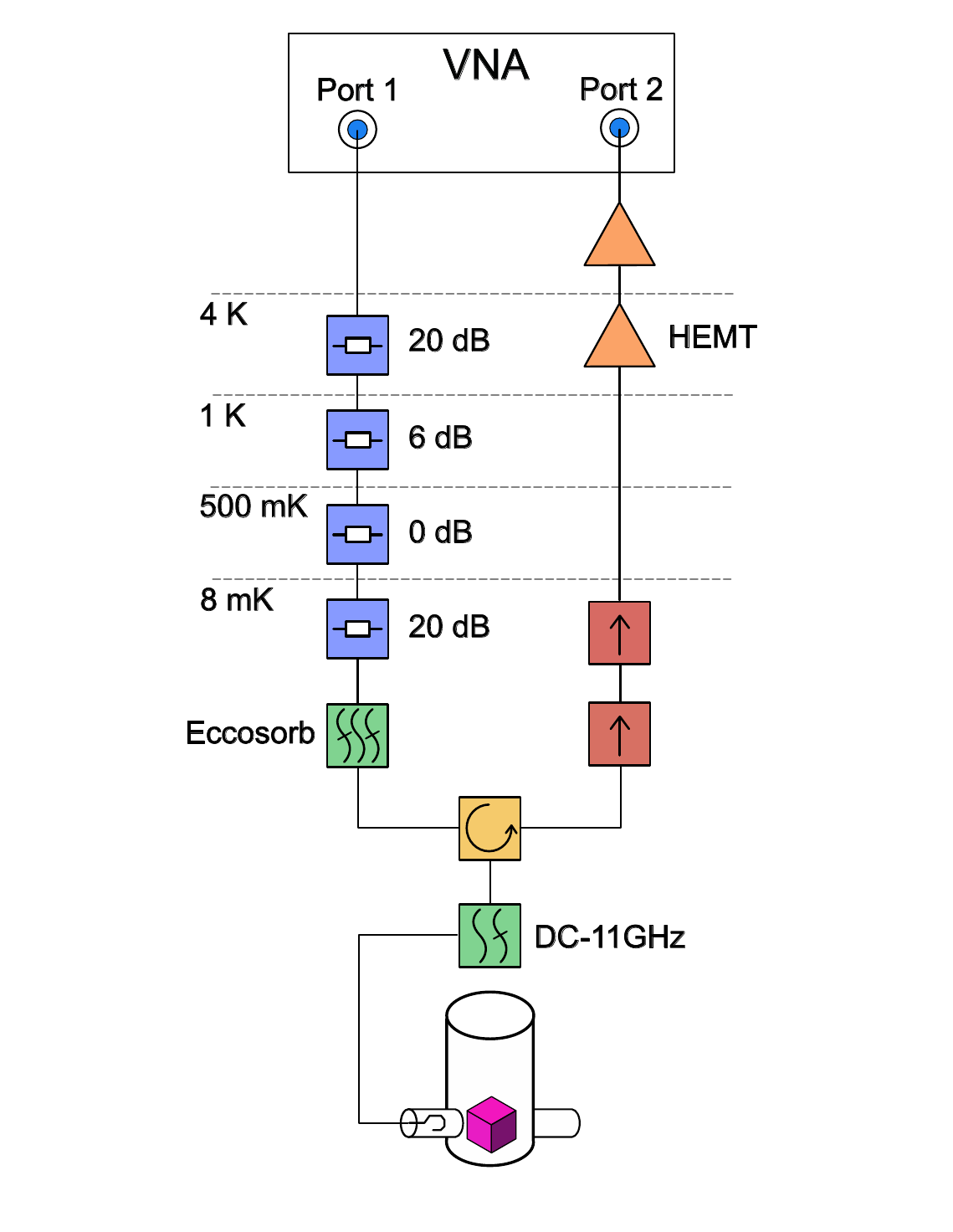}
        \label{fig:ReflectionSetup}
    \end{subfigure}
    \\
    \vspace{-0.5cm}
    \begin{subfigure}[t]{0.48\linewidth}
        \centering
        \subcaption{\raisebox{0ex}[0pt][0pt]{\hspace{200pt}}}
        \includegraphics[scale = 0.35]{Figures/TM010_group_picture.png}
        \label{fig:TM010distro}
    \end{subfigure}
    \quad
    \begin{subfigure}[t]{0.48\linewidth}
        \centering
        \subcaption{\raisebox{0ex}[0pt][0pt]{\hspace{200pt}}}
        \includegraphics[scale = 0.35]{Figures/TE011_group_picture.png}
        \label{fig:TE011distro}
    \end{subfigure}
    \caption{(a) CAD section of the cavity architecture in a one-port measurement configuration. The $z$-cut samples are
     aligned with the main cylinder's axis so that the TM mode excites the $\varepsilon_{\parallel}$ component of the dielectric 
     tensor and the TE mode the $\varepsilon_{\perp}$ component. (b) One-port reflection setup wire diagram, showing the standard implementation of isolators and circulators to prevent noise leakage from the amplifier stages to reach back to the cavity. It is noted that the low-pass DC-11\hspace{1pt}GHz filter was removed for the AlN TE$_{011}$ mode, as it had a resonance frequency above 15 GHz~\cite{Braumüller2018Quantum}. (c) Field distribution vector heatmap of the $\mathrm{TM_{010}}$ and (d) $\mathrm{TE_{011}}$ modes in the cavity in presence of 
       one of the dielectric cubic samples under test.}
\end{figure}

In this study, we investigated a single-crystal, $z$-cut AlN specimen in the form of a cube with a side length of 6~mm, as well as a single-crystal, $z$-cut LiNbO$_{3}$ specimen, likewise prepared as a cube with a side length of 6~mm. The AlN crystal was diamond-cut and did not undergo any polishing\footnote{This is due to limitations of the supplier's manufacturing process.}, while the LiNbO$_{3}$ crystal had its $z$ faces chemically polished. The characterization of the samples was performed by placing them in a custom-designed superconducting cavity to test anisotropic dielectric materials~\cite{zorzetti2023millikelvin}. The cavity (Fig.~\ref{fig:CavityCADSection}) shape comprises a main cylindrical body and four smaller cylindrical side ports positioned perpendicularly to the main volume. The transverse ports allow for aligning the sample in the cavity, as well as installing up to four antennas to couple the microwave modes with the input/output lines. Each sample is placed at the bottom of the main cylinder and held in place by a spring-loaded high-quality single-crystal sapphire rod to compensate for thermal contraction as the temperature decreases. Two superconducting cavities of the above mentioned design were made at Fermilab's machine shop. 
In this experiment, the aluminum cavity hosts the AlN sample and the niobium cavity houses the LiNbO$_{3}$ crystal. Both cavities were pre-processed with chemical treatments to ensure that their intrinsic quality factors are considerably higher than those of the crystal samples (see \emph{Supplementary Note~1}).

A cylindrical cavity is chosen due to the $\mathbf{E}$ field orientation of its eigenmodes: transverse-magnetic (TM) and transverse-electric (TE) modes are well distinguishable, as the electric field aligns either parallel or perpendicular to the cylinder axis (as in Fig.~\ref{fig:TM010distro} and Fig.~\ref{fig:TE011distro}, respectively). Therefore, by exciting the cavity at different frequencies, the crystals' response to the respective electric field orientations can be analyzed, thereby characterizing their full dielectric tensor. For the two materials under study, the tensor is of the form:

\begin{equation}
    \label{eq:anisoEps}
    \varepsilon = \begin{pmatrix}
        \varepsilon_{\perp} & 0 & 0 \\
        0 & \varepsilon_{\perp} & 0 \\
        0 & 0 & \varepsilon_{\parallel}
    \end{pmatrix}.
\end{equation}
with all nonzero entries being complex numbers, that is, $\varepsilon_{m} = \varepsilon_{m}^{\prime} - i\varepsilon_{m}^{\prime\prime} = \varepsilon_{m}^{\prime}\left(1-i\tan{\delta_{m}}\right)$ for $m = \perp, \parallel$, often referred to as the \textit{in-plane} and \textit{out-of-plane} components, respectively.

The cavity is placed in a dilution refrigerator (DR) and the system is cooled down to 10~mK. Through the presence of an antenna positioned in one of the side ports, the cavity is connected to a vector network analyzer (VNA) (Fig.~\ref{fig:ReflectionSetup}). The analyzer is used to perform reflection measurements of the $\mathrm{S}_{11}$ scattering parameter. A fit routine is performed to extract the values of the loaded quality factor $Q_{l}$, the unloaded (internal) quality factor $Q_{0}$, the external quality factor $Q_{ext}$, and the resonance frequency $\nu_{0}$ through the following relation, as in~\cite{probst2015efficient}:
\begin{equation}
    \label{eq:S11Refl}
    S_{11}\left(\nu\right) = ae^{i\alpha}e^{-2\pi i \nu \tau}\left(1-\frac{2\frac{Q_{l}}{\left\|Q_{ext}\right\|}e^{i\phi}}{1 + 2iQ_{l}\left(\frac{\nu - \nu_{0}}{\nu_{0}}\right)}\right),
\end{equation}
where the expression in brackets is a translated circle in the complex impedance plane by the effect of $Q_{ext} \in \mathbb{C}$ and the prefactor $ae^{i\alpha}e^{-2\pi i \nu \tau}$ 
accounts for the presence of the linear transmission line and cable attenuation. The dielectric loss tangent is related to $Q_0$ by taking into account the energy participation ratio (see section \emph{Methods}).


\subsection*{Power sweep measurements}
Measurements were performed twice for all resonant modes under study, during two separate DR cool-down runs. This is done to monitor changes in $Q_{0}$ due to thermal stress and any eventual misalignment of the sample.

A quantitative analysis is conducted on the dataset employing a regression model that comprises two components: a quality factor expressed as a function of power exhibiting TLS behavior \cite{wollack2021loss,crowley2023disentangling}, in conjunction with non-TLS power-insensitive contributions modeled by $Q_\mathrm{other}$. An additional logarithmic component is included in the model for power level above a critical power $P_{c,\log}$ and the logarithmic channel of dissipation is described by the quality factor $Q_{\log}$. The full model reads: 

\begin{subequations}\label{eq:Fx}
\begin{empheq}[ left={\displaystyle\frac{1}{Q_{0}\left(P\right)}=\empheqlbrace}]{align}    
 & \frac{1}{Q_{\mathrm{TLS}}\left(P\right)}+\frac{1}{Q_{\mathrm{other}}} \text{\hspace{0.2cm} if    } P<P_{c,\log}, \label{eq:Q0vsPTLSsaturation} \\
& \frac{1}{Q_{\log}\log{\left(\frac{P}{P_{c,\log}}\right)}} \text{\hspace{0.7cm} if    } P\ge P_{c,\log}. \label{eq:QTLSvsPLog}
\end{empheq}
\end{subequations}
The term associated with TLS is written explicitly as: 

\begin{equation}
    \label{eq:QTLSvsPSqrt}
    Q_{\mathrm{TLS}}\left(P\right) = Q_{\mathrm{TLS,0}}\frac{\sqrt{1+\left(\frac{P}{P_{c}}\right)^{\beta}\tanh{\left(\frac{h\nu}{2k_{B}T}\right)}}}{\tanh{\left(\frac{h\nu}{2k_{B}T}\right)}},
\end{equation}
where $Q_{\mathrm{TLS,0}}$ is the quality factor due to the TLS ensemble at zero temperature and single-photon power, $P_{c}$ is the critical power for TLS saturation, $\nu$ is the resonance frequency of the mode, $T$ is the sample's temperature (assumed equal to the cavity) and $\beta$ is an exponent accounting for non-homogeneous field distribution within the sample.

\begin{figure}[ht!]
    \centering
    \begin{subfigure}[t]{0.48 \linewidth}
        \centering
        \subcaption{\raisebox{0ex}[0pt][0pt]{\hspace{200pt}}}
        \includegraphics[width = 1\linewidth]{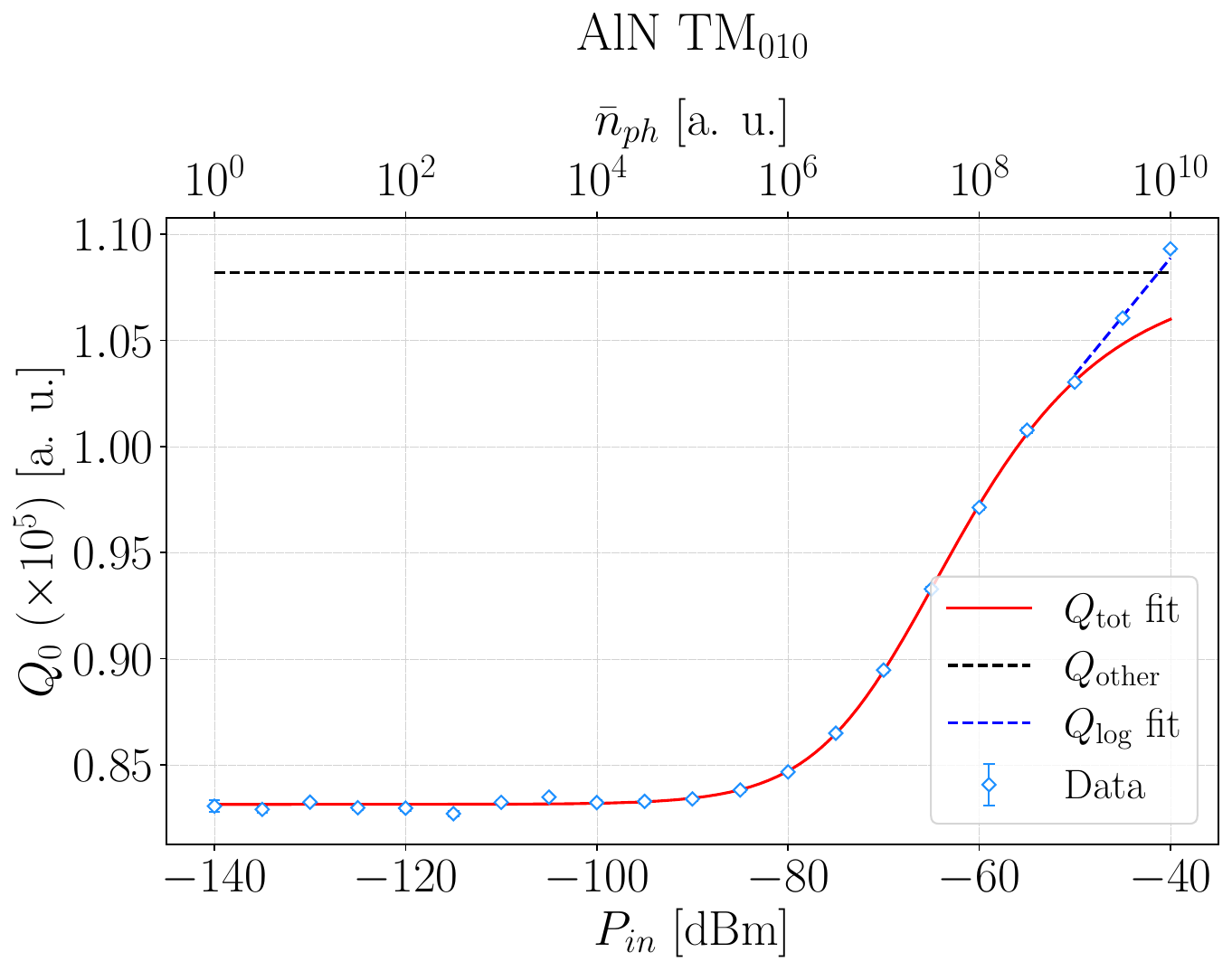}
        \label{fig:Q0AlNPowerSweepFitPiecewiseTM010}
    \end{subfigure}
    \quad
    \begin{subfigure}[t]{0.48 \linewidth}
        \centering
        \subcaption{\raisebox{0ex}[0pt][0pt]{\hspace{200pt}}}
        \includegraphics[width = 1\linewidth]{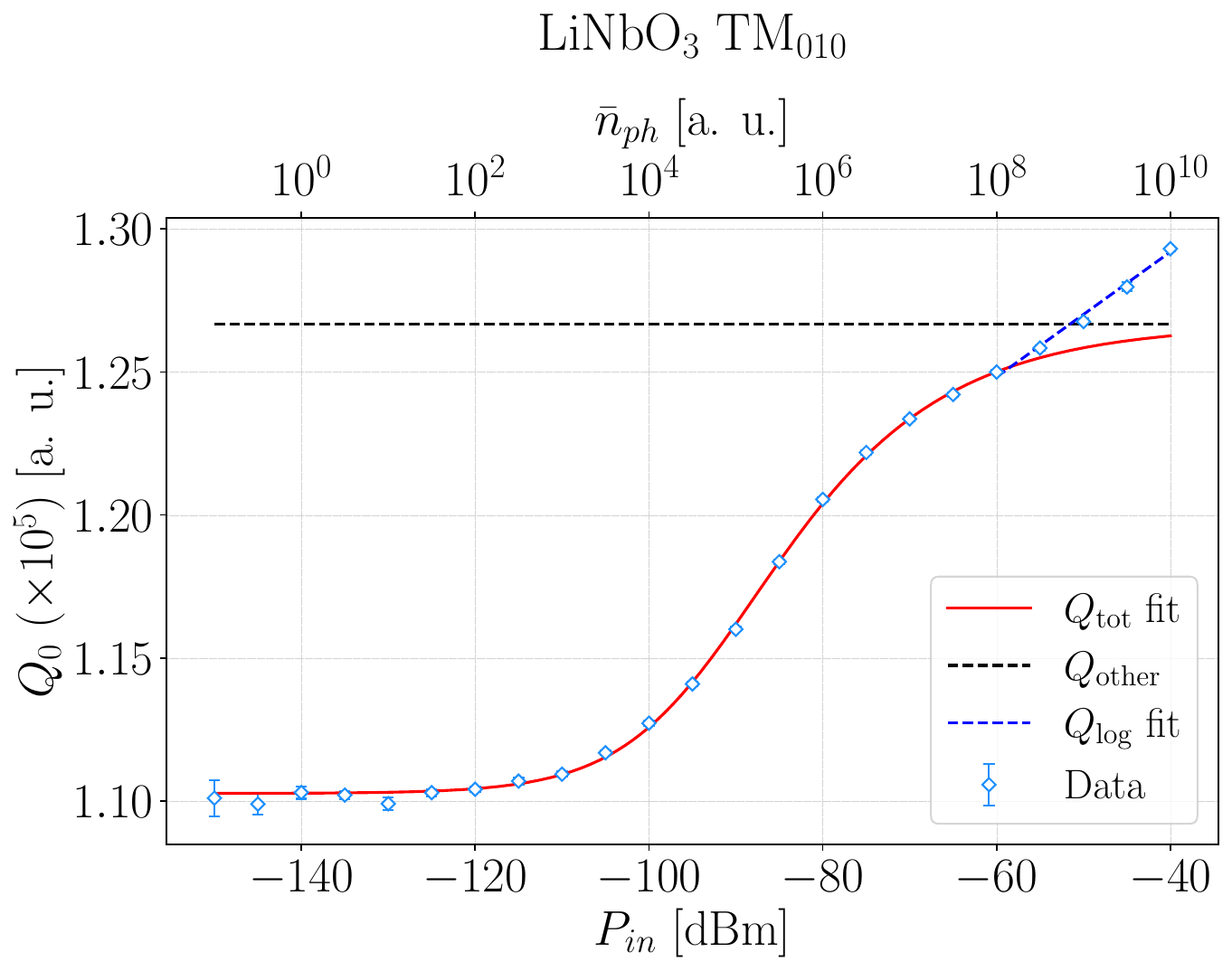}
        \label{fig:Q0LiNbO3PowerSweepFitPiecewiseTM010}
    \end{subfigure}
    \\
    \vspace{-0.3cm}
    \begin{subfigure}[t]{0.48 \linewidth}
        \centering
        \subcaption{\raisebox{0ex}[0pt][0pt]{\hspace{200pt}}}
        \includegraphics[width = 1\linewidth]{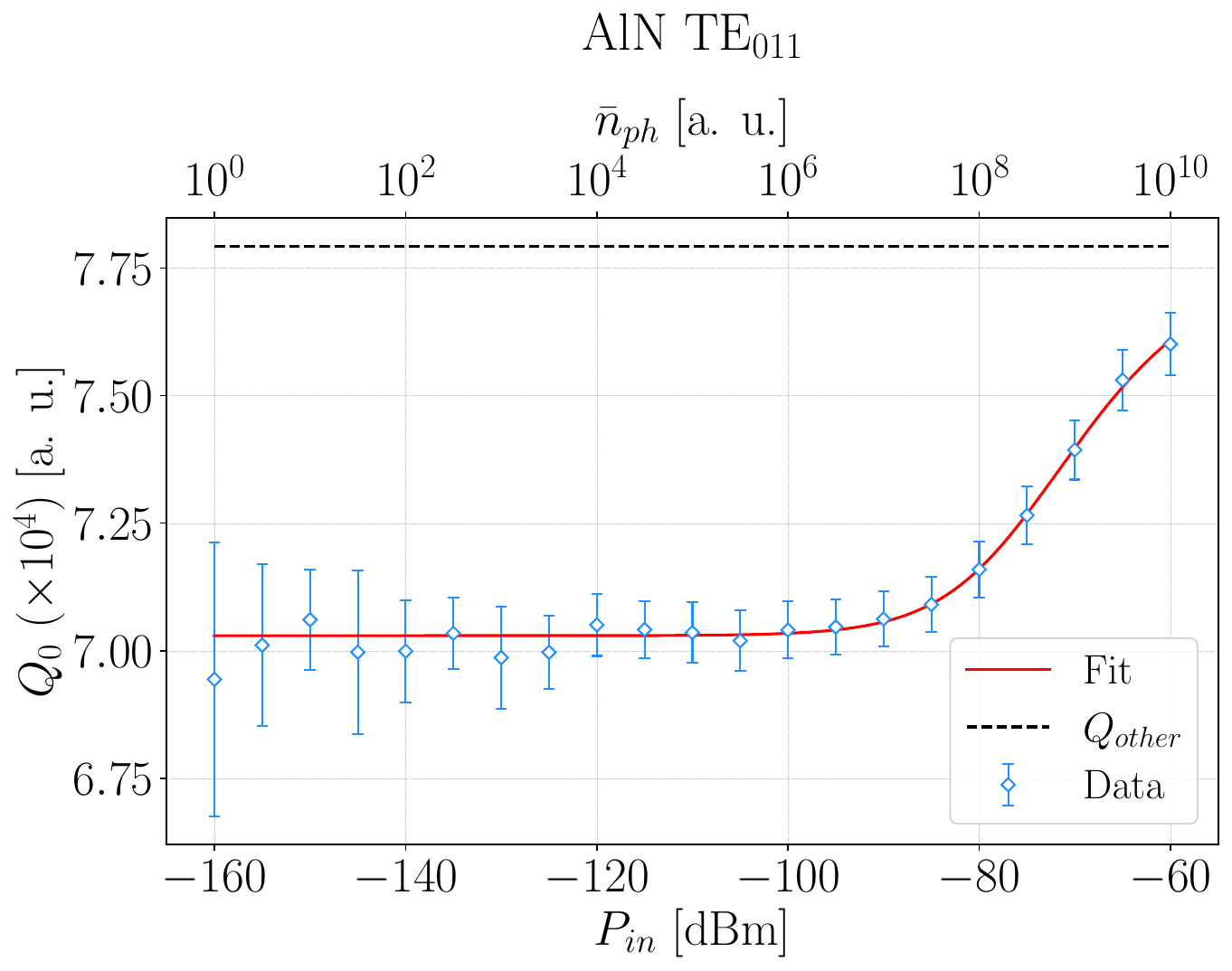}
        \label{fig:Q0AlNPowerSweepFitTLSTE011}
    \end{subfigure}
    \quad
    \begin{subfigure}[t]{0.48 \linewidth}
        \centering
        \subcaption{\raisebox{0ex}[0pt][0pt]{\hspace{200pt}}}
        \includegraphics[width = 1\linewidth]{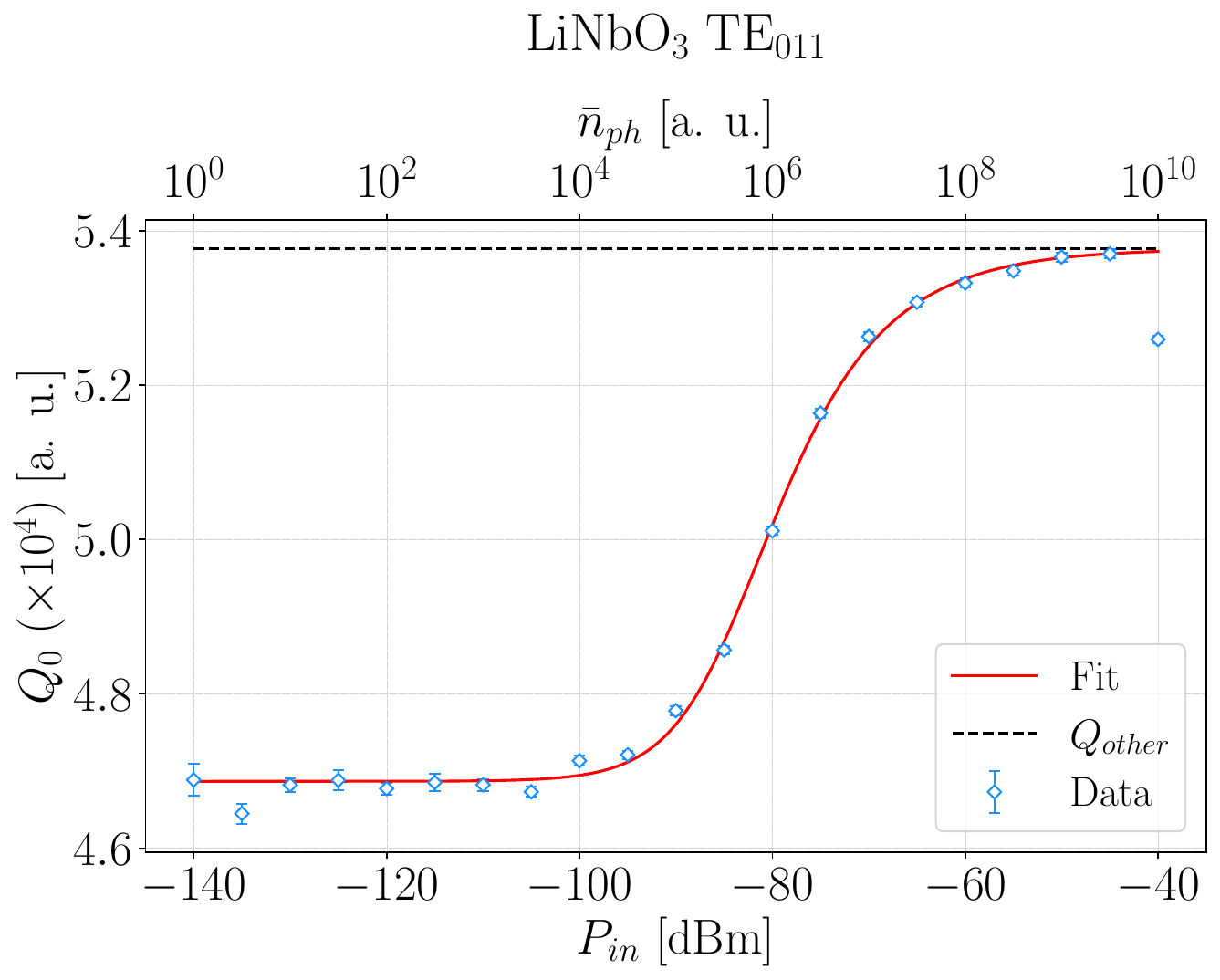}
        \label{fig:Q0LiNbO3PowerSweepFitTLSTE011}
    \end{subfigure}
    \caption{Power sweeps at base temperature of AlN and LiNbO$_{3}$. (a) TM$_{010}$ of AlN with TLS-logarithmic model fit--Eq.~(\ref{eq:QTLSvsPSqrt}) and (\ref{eq:QTLSvsPLog}). (b) TM$_{010}$ of LiNbO$_{3}$ with TLS-logarithmic model fit--Eq.~(\ref{eq:QTLSvsPSqrt}) and (\ref{eq:QTLSvsPLog}). (c) 
    TE$_{011}$ of AlN with TLS model fit, as in Eq.~(\ref{eq:QTLSvsPSqrt}). (d) TE$_{011}$ of LiNbO$_{3}$ with TLS model fit, as in Eq.~(\ref{eq:QTLSvsPSqrt}). The outlier is due to the HEMT gain compression, not fully mitigated by the decompression strategy implemented in the circle fit routine (see \emph{Supplementary Note 2}). Error bars are related to regression to the circle fit model (Eq.~(\ref{eq:S11Refl})) and depend on the signal-to-noise ratio of each $S_{11}$ trace.}
    \label{fig:PowerSweepResultsQ0AlNLiNbO3TM010TE011}
\end{figure}
First, we analyze the $\mathrm{TM}_{010}$ mode through power sweep measurements to probe the materials' properties along the $\varepsilon_{\parallel}$ direction, parallel to the cavity cylinder axis. Overall, the two materials behave similarly: the quality factor is constant in power up to a certain value, above which it starts to increase, resulting in a TLS-like response, until logarithmic behavior is observed in  $Q_{0}$ vs. $P_{in}$  above a second threshold power value (Fig.~\ref{fig:PowerSweepResultsQ0AlNLiNbO3TM010TE011}a, b). Regression results are shown in Table~\ref{tab:PwrSweepRegrResultsTM010}.

We then analyze the TE$_{011}$ mode to probe microwave properties of the materials in the $\varepsilon_{\perp}$ direction, perpendicular to the cavity cylinder axis (Fig.~\ref{fig:PowerSweepResultsQ0AlNLiNbO3TM010TE011}c, d). It is noted that, from FEM simulations (see \emph{Supplementary Note 1}), the AlN TE$_{011}$ mode frequency is located around 15.5~GHz at room temperature, which falls outside the working bandwidth of the circulator and isolators available in this setup. However, the signal-to-noise ratio provided by the HEMT amplifier is still enough to extract all information from the resonance. On the other hand, the LiNbO$_{3}$ mode is within the bandwidth of all components, about 7~GHz. For these resonances, the quality factor as a function of power does not show a logarithmic response for a high electric field magnitude\footnote{In the case of AlN, the line attenuation is substantially greater due to its elevated frequency, resulting in the maximum power delivered to the cavity being limited to -65 dBm. This level of power is outside the range where the unusual logarithmic response is observed in TM$_{010}$ resonances.}. Therefore, we fit the entire trace with Eq.~(\ref{eq:Q0vsPTLSsaturation}). Regression results are shown in Table~\ref{tab:PwrSweepRegrResultsTE011}.

\input{Tables/Power_sweep_regression_results_TM010}

\input{Tables/Power_sweep_regression_results_TE011}

The observed TLS behavior across all resonances for intermediate- and high-power intervals is peculiar compared with that of other dielectric materials in the same frequency range. In particular, studies conducted on dielectric substrates and thin layers typically do not show saturation to the $Q_{\mathrm{other}}$ component of the model at high power \cite{crowley2023disentangling,gao2008thephysics}. Here, in the case of bulk single crystals, the quality factor rapidly converges to the second term of Eq.~(\ref{eq:Q0vsPTLSsaturation}), as shown in Fig.~\ref{fig:PowerSweepResultsQ0AlNLiNbO3TM010TE011} and is particularly evident for the TE$_{011}$ mode (Fig.~\ref{fig:PowerSweepResultsQ0AlNLiNbO3TM010TE011}d). 

From the theory of TLS, only a small number of TLS can absorb energy from the microwave field in the cavity, i.e., those with energy splitting $\zeta \simeq h\nu$, where $\nu$ is the field's resonance frequency. This explains the observed saturation toward $Q_{\mathrm{other}}$ at a relatively low average photon population in the cavity~\cite{muller2019towards}.
In the low temperature and low power regime, TLS with energy on the order of 10~\textmu eV are typically linked to tunneling atoms, collective motion of small atomic domains inside amorphous solids or so-called dangling bonds~\cite{muller2019towards}. Given the high degree of regularity and purity of the single crystals under test, these physical entities are located on the outer surfaces of the samples, where a thin layer of oxide or nonstoichiometric material is present. Room-temperature analysis of the crystals' $z$ surfaces via XPS and time-of-flight SIMS corroborate these hypotheses (see subsection \emph{Room-temperature material analysis}). 

Interestingly, both materials show a stark difference between in-plane and out-of-plane dielectric responses. While $Q_{0}$ reaches effective saturation to $Q_{\mathrm{other}}$ in the TE$_{011}$ mode\footnote{with the previous caveats on the AlN resonance.} (Fig.~\ref{fig:PowerSweepResultsQ0AlNLiNbO3TM010TE011}c, d), in the TM$_{010}$ resonances a residual logarithmic increase of $Q_{0}$ is observed (Fig.~\ref{fig:PowerSweepResultsQ0AlNLiNbO3TM010TE011}a, b). The unusual logarithmic behavior is consistent with the presence of strongly interacting TLS, excited by the transverse magnetic resonant mode and described by the general tunneling model (GTM)~\cite{Faoro2012Internal,Burnett2016Analysis}. The nature of the behavior of the materials in this regime is elusive, and it is possible that the observed effects originate from additional degrees of freedom in the material, such as piezoelectricity or ferroelectricity, which express themselves as strongly interacting TLS. In this case, these strongly interacting TLS may not be confined to the sample surfaces but could instead be distributed throughout the crystal bulk or associated with collective lattice motion driven by piezoelectric or ferroelectric mechanisms. The difference in behavior between TM$_{010}$ and TE$_{011}$ mode responses at high power is also linked to the clamping strategy used to keep the crystals in place: the sapphire rod and bottom of the cavity cylinder create a \emph{phononic trap} that enhances some bulk acoustic properties in the $z$-direction, perpendicular to the boundary surfaces. This, ultimately, results in accentuated high electric power response when exciting the TM$_{010}$ mode of the crystal~\cite{goryachev2014effects,parashar2024upconversion}. On the other hand, there is no mechanical constraint on the transversal portion of the crystals, resulting in very weak phononic confinement. Further analysis of these phenomena is needed to understand why the quality factor of TM$_{010}$ modes increases as the piezoelectric contribution becomes non-negligible, as well as a detailed simulation of the clamp-induced phononic trap.

\subsection*{Temperature sweep measurements}
Evidence of TLS is also found by sweeping the temperature. The materials' response as a function of temperature is analyzed by parameterizing the samples' losses with a conventional three-component model \cite{crowley2023disentangling,gao2008thephysics}: 
\begin{equation}
    \label{eq:QTotVsTP}
    \frac{1}{Q_0\left(\bar{n},T\right)} = \frac{1}{Q_{\mathrm{TLS}}\left(\bar{n}, T\right)} + \frac{1}{Q_{\mathrm{QP}}\left(T\right)} + \frac{1}{Q_{\mathrm{other}}}.
\end{equation}
The form of the TLS-related term also includes an explicit dependence of critical power vs. temperature:

\begin{equation}
    \label{eq:QTLSvsPandT}
    Q_{\mathrm{TLS}}\left(\bar{n},T\right)=Q_{\mathrm{TLS,0}}\frac{\sqrt{1+\left(\frac{\bar{n}^{\beta_{2}}}{DT^{\beta_{1}}}\right)\tanh{\left(\frac{h\nu}{2k_{B}T}\right)}}}{\tanh{\left(\frac{h\nu}{2k_{B}T}\right)}},
\end{equation}
with a dedicated exponent $\beta_{1}$ for the temperature dependence of the critical power $P_{c}=DT^{\beta_1}$ that gauges how much it deviates from a linear dependence and a second exponent $\beta_{2}$ that controls the input power dependence and accounts, once again, for field non-homogeneity inside the sample. The factor $D$ is related to the amplitude of the electric field of the TLS critical power. As a reminder, $\bar{n}\left(P\right)$ is the average intracavity photon number, function of the input power through Eq.~(\ref{eq:AvgPhNumb}).

The second term in Eq.~(\ref{eq:QTotVsTP}) models the losses due to quasiparticles on the cavity walls from the breaking of Cooper pairs:

\begin{equation}
    \label{eq:QqpvsT}
    Q_{\mathrm{QP}}\left(T\right) = A_{\mathrm{QP}}\frac{e^{\frac{\Delta_{0}}{k_{B}T}}}{\sinh\left(\frac{h\nu}{2k_{B}T}\right)K_{0}\left(\frac{h\nu}{2k_{B}T}\right)},
\end{equation}
where $A_{\mathrm{QP}}$ is a magnitude proportional to the kinetic inductance of the quasiparticles on the inner walls of the cavity, $\Delta_{0}~=~\frac{e^{-\gamma}}{\pi}k_{B}T_{c}$ is the superconducting gap with $\gamma$ as the Euler-Mascheroni constant, $T_{c}$ is the critical temperature of the superconductor, $K_{0}$ is the modified Bessel function of the second kind, and $\nu$ is the resonance frequency of the mode under test \cite{crowley2023disentangling,gao2008thephysics}. A power and temperature independent factor $Q_{\mathrm{other}}$ is added to Eq.~(\ref{eq:QTotVsTP})  to account for any additional sources of losses. The fitting parameters used to implement the regression are six: $Q_{\mathrm{TLS,0}}$, $D$, $\beta_{1}$, $A_{\mathrm{QP}}$, $T_{c}$, and $Q_{\mathrm{other}}$; $\beta_{2}$ is obtained from the previous base-temperature power regressions \cite{gao2008thephysics,gao2008equivalence}. The TLS model also predicts a temperature dependence of the resonance frequency~\cite{muller2019towards,gao2008thephysics,Faoro2012Internal,gao2008equivalence}. These analyses are discussed in \emph{Supplementary Note 4}.

\begin{figure}[htbp]
    \centering
\includegraphics[width=1\linewidth]{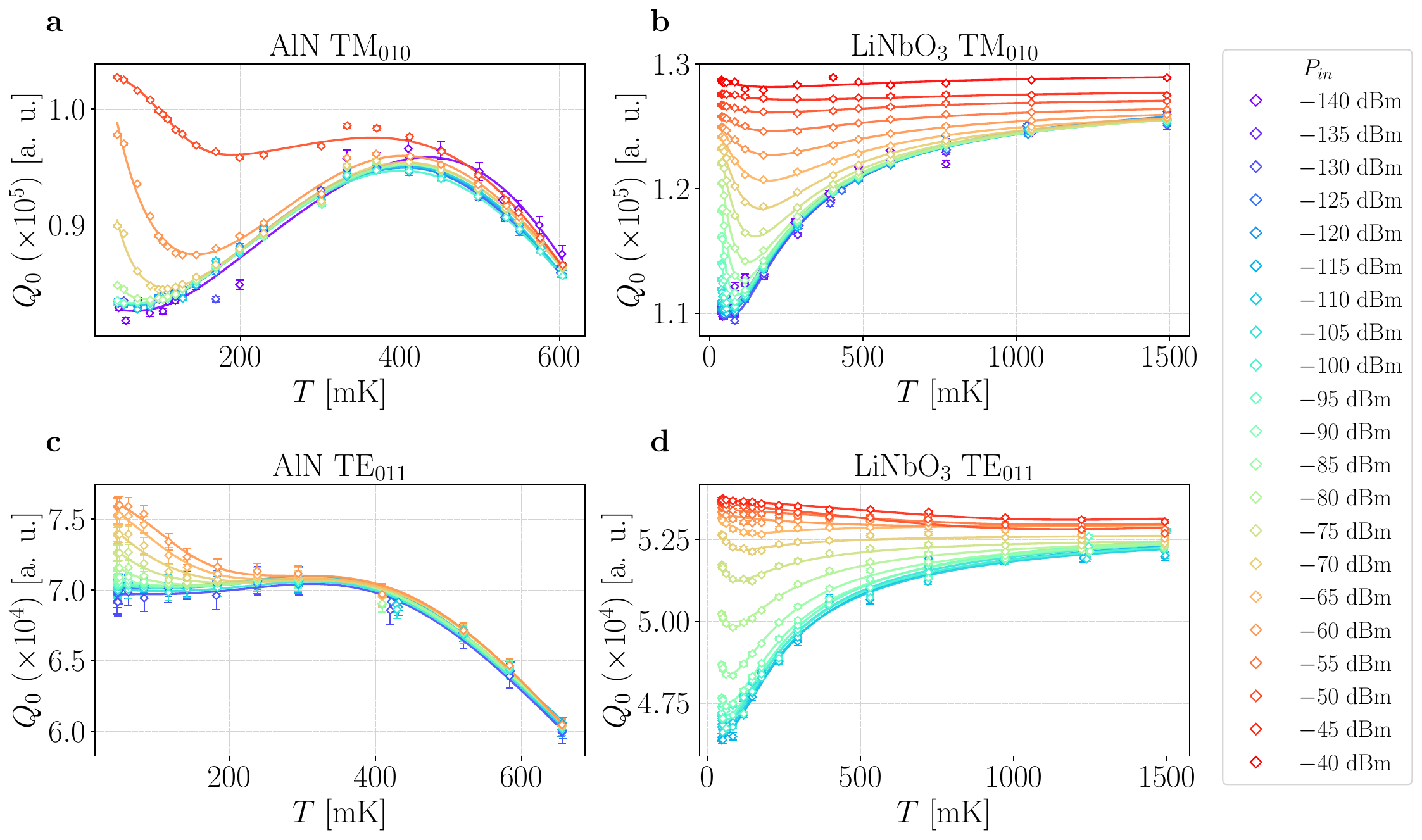}
    \caption{Samples TM$_{010}$ (a)(b) and TE$_{011}$ (c)(d) mode internal quality factor versus temperature measurements at different cavity input power and corresponding regressions to model in Eq.~(\ref{eq:QTotVsTP}).
    In the case of AlN, the $Q_{0}\left(\bar{n},T\right)$ curves show a distinct local/global maximum for the transverse magnetic resonance and an inflection point for the transverse electric one around $T\simeq400$ mK, indicating a clear presence of non-negligible quasiparticle dissipation in the aluminum cavity walls at high temperatures. For LiNbO$_{3}$, the absence of quasiparticle-related contribution in the $Q_{0}\left(\bar{n},T\right)$ data series is consistent with the high critical temperature of the niobium cavity. The difference in high power behavior between the in-plane and out-of-plane dielectric responses observed in base-temperature power sweeps is retrieved here at any temperature, indicating that the logarithmic behavior in Eq.~(\ref{eq:QTLSvsPLog}) observed for the transverse magnetic resonance is temperature-independent. Error bars are, once again, related to regression to the circle fit model (Eq.~(\ref{eq:S11Refl})) and depend on the signal-to-noise ratio of each $S_{11}$ trace.}
    \label{fig:Q0vsTAllModesAllSamples}
\end{figure}

For the AlN sample, the temperature dependence of the quality factor follows the full model in Eq.~(\ref{eq:QTotVsTP}), showing a good agreement on both dielectric tensor directions. A TLS-driven behavior appears at low temperatures, with a local or global maximum at around 400~mK for the TM$_{010}$ mode, depending on the input power of the cavity (Fig.~\ref{fig:Q0vsTAllModesAllSamples}b), and a horizontal inflection point for the TE$_{011}$ mode (Fig.~\ref{fig:Q0vsTAllModesAllSamples}d). At higher temperatures, the quality factor decreases due to Cooper pair breaking and quasiparticle formation in the aluminum walls of the cavity. From the regression to the three-component model of both resonances, we obtain a critical temperature for the cavity of $T_{c} = 1.3 \ \pm \ 0.2 $ K, in agreement with the superconducting critical temperature of bulk aluminum, of 1.2 K~\cite{kittel2005introduction}. The same regression yields a zero-temperature TLS quality factor of $Q^\mathrm{TM010}_{\mathrm{TLS,0}}= \left(3.4 \ \pm \ 0.1\right) \times 10^{5}$ at low power for the transverse magnetic resonance and of $Q^\mathrm{TE011}_{\mathrm{TLS,0}}= \left(7.4 \ \pm \ 0.2\right) \times 10^{5}$ for the transverse electric one, compatible with the values extrapolated by the base temperature power sweep regression in Tables~\ref{tab:PwrSweepRegrResultsTM010}~-~\ref{tab:PwrSweepRegrResultsTE011}. At higher power (above -60~dBm), the zero-temperature TLS quality factor component estimation from the fit also increases, up to $10^{6}$ in both cases. This power dependence of $Q_{\mathrm{TLS,0}}$ estimation has been previously observed in literature \cite{leon2021materials} and it is commonly attributed to limitations of the regression model in Eq.~(\ref{eq:QTLSvsPandT}): at higher power, the internal quality factor tends to saturate. In the specific case of the materials under study, the increase in $Q_{\mathrm{TLS,0}}$ could also indicate additional degrees of freedom, as pointed out in the analysis of the power sweep measurements, potentially related to strongly interacting TLS. The values of power-independent quality factors,  $Q^\mathrm{TM010}_{\mathrm{other}}= \left(1.18 \ \pm \ 0.07\right)\times10^{5}$ for the transverse magnetic mode and $Q^\mathrm{TE011}_{\mathrm{other}}= \left(7.7 \ \pm \ 0.3\right)\times10^{4}$ for the transverse electric mode, are compatible with the results of the base-temperature power sweep fits in Tables~\ref{tab:PwrSweepRegrResultsTM010}~-~\ref{tab:PwrSweepRegrResultsTE011}.

In evaluating the internal quality factor of the LiNbO$_{3}$ resonances, we notice that the loss contribution due to quasiparticles on the cavity walls is absent in the explored temperature range. This is consistent with the higher critical temperature value of niobium and the typical onset of quasi-particle-related losses, which occur at $T\gtrsim \frac{T_{c}}{4}$. Thus, the regressions to the model are done by omitting the quasiparticle contribution. The modified two-component model agrees very well with the data series for any given cavity input power (Fig.~\ref{fig:Q0vsTAllModesAllSamples}). From the fits, we extract a zero-temperature TLS quality factor of $Q^\mathrm{TM010}_{\mathrm{TLS,0}}= \left(8.0 \ \pm \ 0.3\right)\times 10^{5}$ for the transverse magnetic mode and $Q^\mathrm{TE011}_{\mathrm{TLS,0}}= \left(3.9 \ \pm \ 0.2\right)\times 10^{5}$ for the transverse electric one at low power, compatible with the value extracted from the base-temperature power sweeps (Tables~\ref{tab:PwrSweepRegrResultsTM010}~-~\ref{tab:PwrSweepRegrResultsTE011}). For excitation power exceeding -60 dBm, we retrieve a high-power TLS quality factor of, again,  $10^{6}$ due to the limitations in Eq.~(\ref{eq:QTLSvsPandT}) at high electric field magnitude previously discussed. From the same regressions, we find a power-independent quality factor value of $Q^\mathrm{TM010}_{\mathrm{other}}= \left(1.28 \ \pm \ 0.01\right)\times10^{5}$ for the transverse magnetic mode and $Q^\mathrm{TE011}_{\mathrm{other}}= \left(5.4 \ \pm \ 0.2\right)\times10^{4}$ for the transverse electric one, constant in power and once again compatible with the ones extrapolated from base-temperature power sweeps. The frequency shift of each resonant mode as a function of temperature is also analyzed for both materials to corroborate the TLS behavior observed in the $Q$ factor versus temperature traces. A detailed analysis of this behavior is reported in \emph{Supplementary Note 4}. 

\begin{figure}[ht!]
    \centering
    \begin{subfigure}[t]{0.4 \linewidth}
        \centering
        \subcaption{\raisebox{0ex}[0pt][0pt]{\hspace{200pt}}}
        \includegraphics[width = 0.85\linewidth]{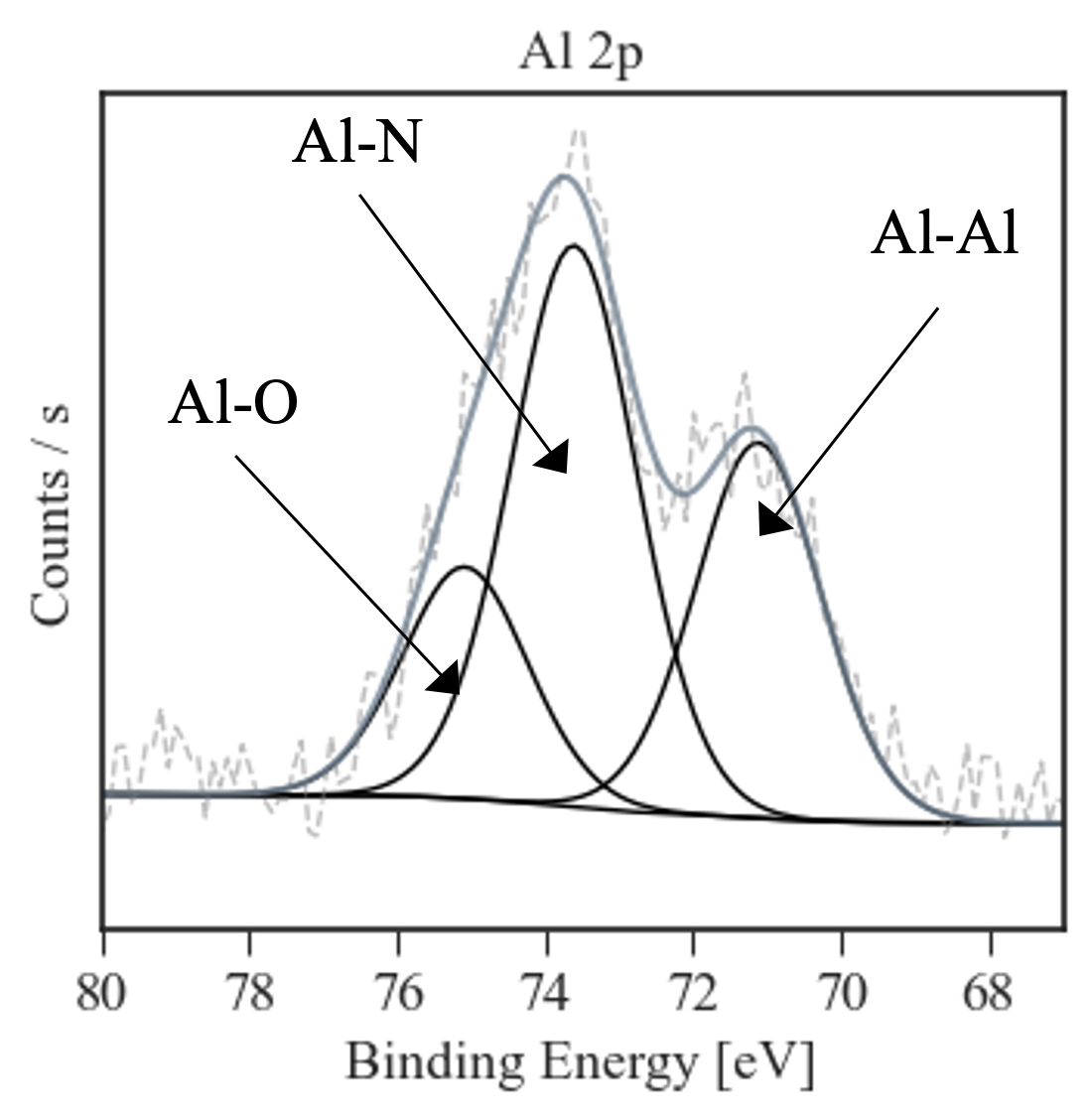}
        \label{fig:XPS Al 2p}
    \end{subfigure}
    \quad
    \begin{subfigure}[t]{0.4 \linewidth}
        \centering
        \subcaption{\raisebox{0ex}[0pt][0pt]{\hspace{200pt}}}
        \includegraphics[width = 0.85\linewidth]{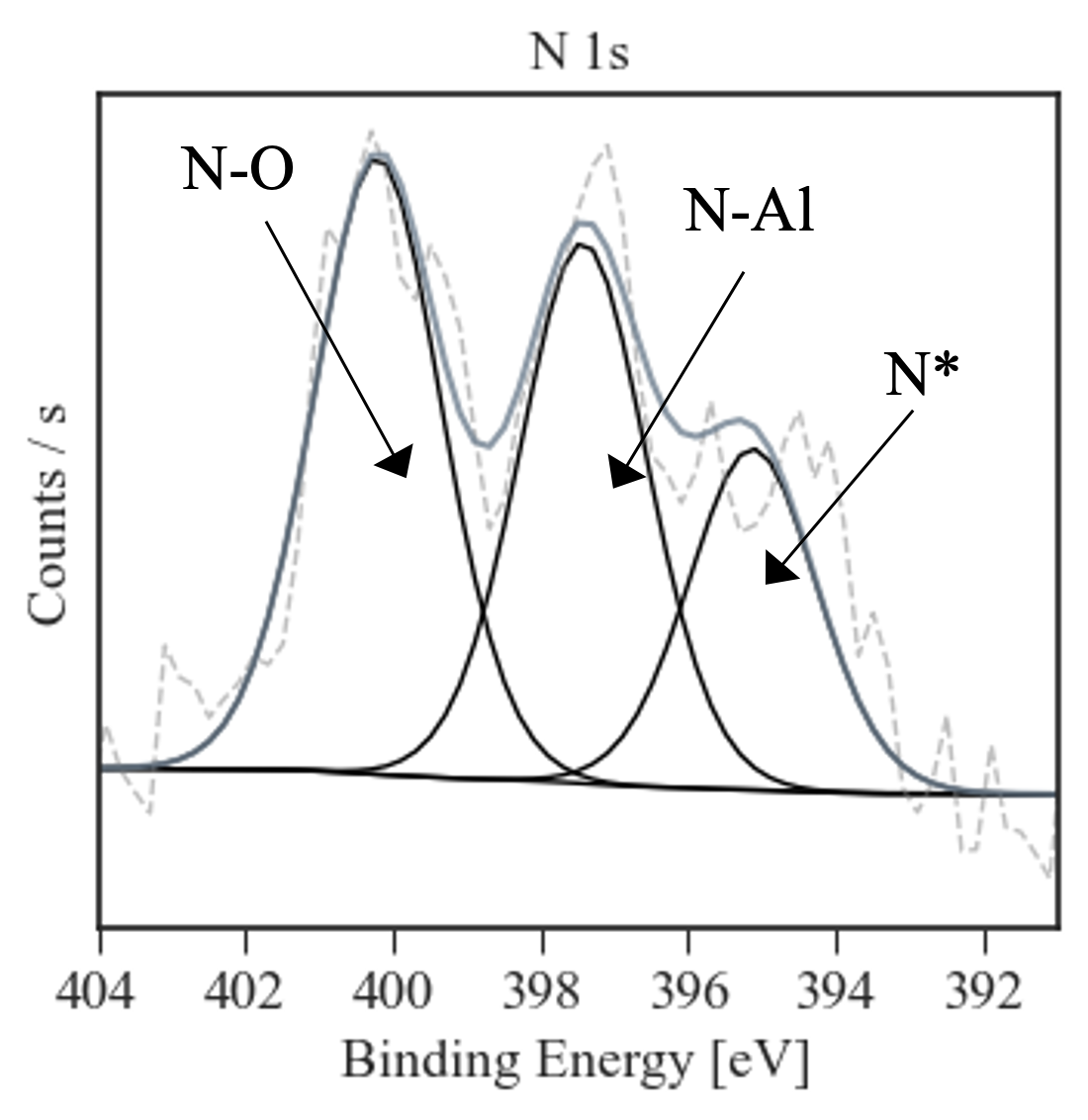}
        \label{fig:XPS N 1s}
    \end{subfigure}
    \\
    \vspace{-0.3cm}
    \begin{subfigure}[t]{0.4 \linewidth}
        \centering
        \subcaption{\raisebox{0ex}[0pt][0pt]{\hspace{200pt}}}
        \includegraphics[width = 0.87\linewidth]{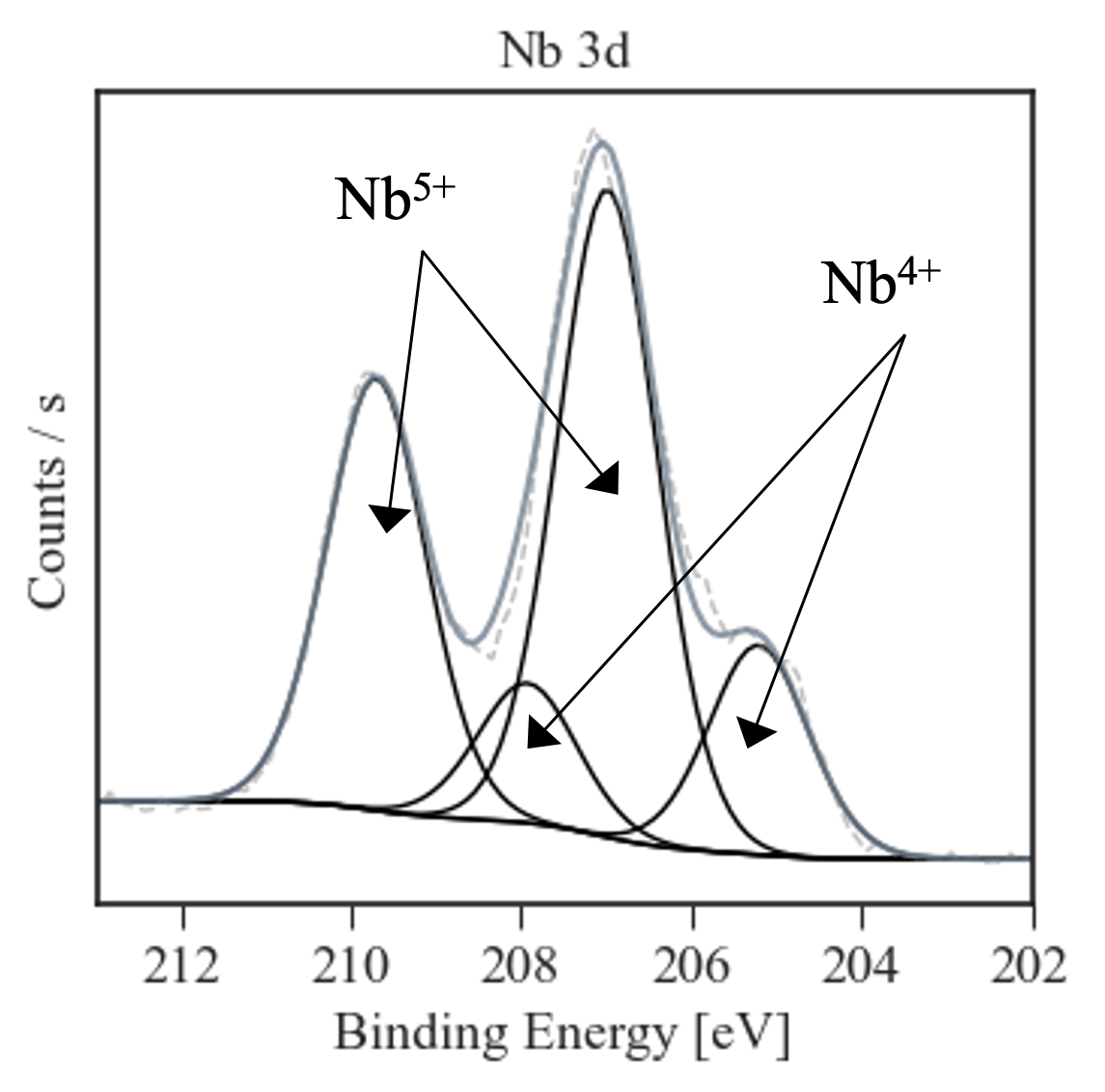}
        \label{fig:XPS Nb 3d}
    \end{subfigure}
    \quad
    \begin{subfigure}[t]{0.4 \linewidth}
        \centering
        \subcaption{\raisebox{0ex}[0pt][0pt]{\hspace{200pt}}}
        \includegraphics[width = 0.85\linewidth]{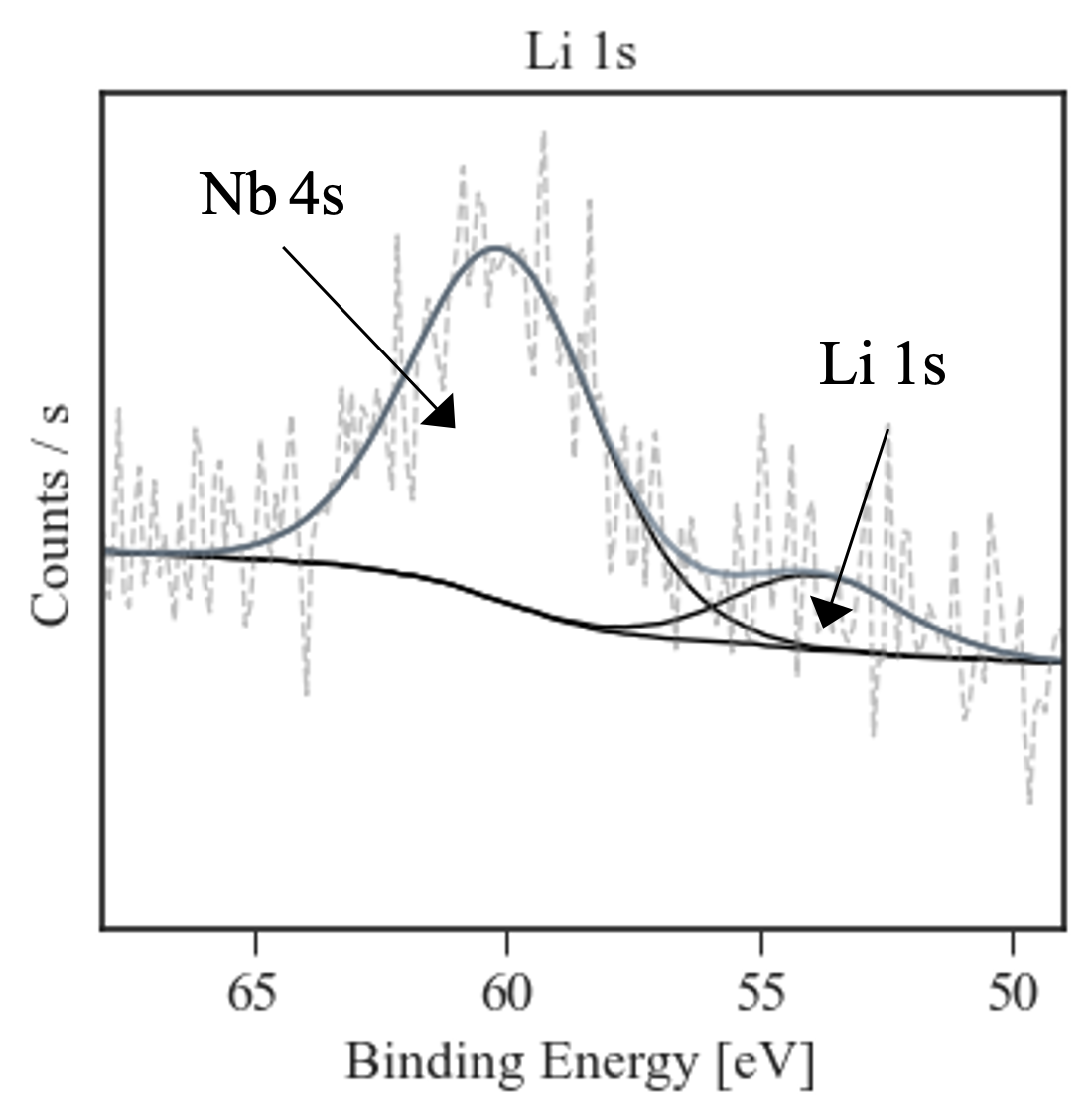}
        \label{fig:XPS Li 1s}
    \end{subfigure}
    \caption{High resolution XPS of (a) Al 2p, and (b) N 1s of the AlN bulk crystal, and (c) Nb 3d and (d) Li 1s of the LiNbO$_{3}$ bulk crystal. The solid and dashed gray lines represent the envelope and the raw spectrum, respectively.}
    \label{fig:Material Characterization}
\end{figure}

\subsection*{Room-temperature material analysis}

To probe the presence of impurities at the surface and within the sample, the crystals were analyzed using X-ray photoelectron spectroscopy (XPS) and time-of-flight secondary ion mass spectrometry (ToF-SIMS). Figures~\ref{fig:XPS Al 2p} and ~\ref{fig:XPS N 1s} show the high-resolution Al 2p and N 1s spectra taken with XPS from the AlN crystal, respectively. 
For AlN, the presence of the Al–Al component in addition to the Al-N indicates a nitrogen-deficient surface composition. The high-resolution N 1s spectrum consists of peaks at 397.42 eV (N–Al), 400.21~eV (N–O), and 395.12~eV (N*). The N–Al peak position agrees with literature values, while the N–O peak appears at a slightly higher binding energy than typical nitrogen oxides, likely due to carbon contamination forming organic nitrogen species \cite{aumann1995oxidation,motamedi2014xps}. The N* peak at 395.12~eV corresponds to nitrogen bound to a complex of multiple aluminum atoms rather than to AlN \cite{chale2019deposit}. In addition, the Al 2p/N 1s atomic ratio of 1.35 further confirms a nitrogen-deficient, aluminum-rich surface. 
Using ToF-SIMS, we find a variety of impurities at the surface of the AlN crystal. Namely, as shown in \emph{Supplementary Note 3}, in addition to Al and N signal, we observe significant impurities, such as O$^-$, C$^-$, H$^-$, F$^-$, Cl$^-$, Si$^-$, P$^-$, S$^-$, Ni$^-$, and Cu$^-$ signals.

In terms of the LiNbO$_{3}$ crystal, from XPS we find that the Nb atoms present in this compound primarily exhibit the 5+ oxidation state, as shown in Fig.~\ref{fig:XPS Nb 3d}.
Oxygen vacancies have been shown to scale with TLS loss in the case of amorphous Nb$_2$O$_5$ \cite{bafia2024oxygen}. From the Li 1s spectra in Fig.~\ref{fig:XPS Li 1s}, the Li/Nb ratio is 1.38 $\pm$ 0.1, indicating an excess of lithium, potentially due to preferential segregation or diffusion during processing \cite{skryleva2016xps}. Using ToF-SIMS, we find that similar to the AlN crystal, the LiNbO$_3$ crystal exhibits appreciable counts of O$^-$, C$^-$, H$^-$, F$^-$, Cl$^-$, Si$^-$, P$^-$, S$^-$, Ni$^-$, and Cu$^-$ signals.

Both crystals exhibit a high degree of surface contamination that is likely introduced during handling and cutting. In addition, the formation of surface oxides with varying stoichiometry in both AlN and LiNbO$_{3}$ is evident. These features, together with stoichiometric deviations such as the nitrogen deficiency in AlN and the excess lithium in LiNbO$_{3}$ may contribute to microwave loss and potentially the TLS behavior observed for these materials.

%% file: Tables/Power_sweep_regression_results_TM010.tex
\setlength{\tabcolsep}{2.5pt} 
\begin{center}
\begin{threeparttable}[htbp!]
    \centering
    \caption{Piecewise fit results of the TM$_{010}$ mode internal quality factor as a function of power data series of the two materials, as in Eq.~(\ref{eq:Q0vsPTLSsaturation}) and (\ref{eq:QTLSvsPLog}).}
    \begin{tabular*}{1\textwidth}{@{}lcccccc@{}}
    \toprule
    Sample & $Q_{\mathrm{TLS,0}}\left(\times 10^{5}\right)$ & $Q_{\mathrm{other}}\left(\times 10^{5}\right)$ & $Q_{0}\left(\bar{n}_{ph}\rightarrow0\right)\left(\times 10^{5}\right)$ & $P_{c}$ [dBm] & $Q_{\log}\left(\times 10^{3}\right)$ & $P_{c,\log}$ [dBm]\\
    \midrule
        AlN\tnote{1} & 3.59 $\pm$ 0.04 &  1.08 $\pm$ 0.04  &  0.829 $\pm$ 0.002  & -70.3 $\pm$ 0.1 & 2.39 $\pm$ 0.07 & -50 $\pm$ 1\\
        LiNbO$_{3}$\tnote{2} & 8.52 $\pm$ 0.09 & 1.27 $\pm$ 0.01 & 1.11 $\pm$ 0.01 & -93.9 $\pm$ 0.3 & 1.01 $\pm$ 0.01 & -60 $\pm$ 1 \\
    \bottomrule
    \end{tabular*}
    \begin{tablenotes}
        \item[1] {Resonant frequency $\nu_{0}=9.5661021(2)$~GHz;}
        \item[2] {Resonant frequency $\nu_{0}=7.6750283(4)$~GHz.}
    \end{tablenotes}
    
    \label{tab:PwrSweepRegrResultsTM010}
\end{threeparttable}
\end{center}

%% file: Tables/Power_sweep_regression_results_TE011.tex
\setlength{\tabcolsep}{4pt} 
\begin{center}
\begin{threeparttable}[htbp!]
    \centering
    \caption{TLS fit results of the TE$_{011}$ mode internal quality factor versus power data series of the two materials.}
    \begin{tabular*}{0.77\textwidth}{@{}lcccc@{}}
    \toprule
    Sample & $Q_{\mathrm{TLS,0}}\left(\times10^5\right)$ & $Q_{\mathrm{other}}\left(\times10^4\right)$ & $Q_{0}\left(\bar{n}_{ph}\rightarrow0\right)\left(\times10^4\right)$ & $P_{c}$ [dBm] \\
    \midrule
        AlN\tnote{1} & 7.2 $\pm$ 0.6 &  7.8 $\pm$ 0.3  &  7.0 $\pm$ 0.4 & -75 $\pm$ 1\\
        LiNbO$_{3}$\tnote{2} & 3.64 $\pm$ 0.03 & 5.38 $\pm$ 0.04  &  4.68 $\pm$ 0.04 & -84.9 $\pm$ 0.2\\
    \bottomrule
    \end{tabular*}
    \begin{tablenotes}
        \item[1]{Resonance frequency $\nu_{0}=$ 15.508453(4) GHz};
        \item[2]{Resonance frequency $\nu_{0}=$ 7.020667(1) GHz.}
    \end{tablenotes}
    
    \label{tab:PwrSweepRegrResultsTE011}
\end{threeparttable}
\end{center}

%% file: methods.tex
\subsection*{Analysis of the dielectric loss tangent}
The dielectric loss tangent is related to $Q_{0}$ via the expression \cite{chuanlan2022resonant}:

\begin{equation}
    \label{eq:Qaniso}
    \frac{1}{Q_{0}} = \sum_{m = \perp, \parallel} p_{m}\tan{\delta}_{m},
\end{equation}
where $p_{m}$ is known as \textit{energy participation ratio} and quantifies the fraction of the electric energy stored in a given dielectric volume $\Xi_{m}$. It is characterized by a certain relative permittivity $\varepsilon_{m}$ with respect to the total energy:

\begin{equation}
    \label{eq:partRatio}
    p_{m} = \frac{\frac{\varepsilon_{m}^{\prime}}{2}\int_{\Xi_{m}}\left\|\mathbf{E}\right\|^{2}d\Xi_{m}}{E_{tot}}.
\end{equation}

Through the use of the relation in Eq.~(\ref{eq:Qaniso}) and considering different orientations of the electric fields in the cavity for the TM and TE modes, we can decouple the dielectric behavior in both directions and determine the response of the crystals for the two orthogonal directions. The corresponding energy participation ratios are computed through finite-element method (FEM) simulations of the cavity, with each crystal sample placed within it, in addition to the sapphire.
The expression of the internal quality factor in Eq.~(\ref{eq:Qaniso}) is valid under the assumption that the losses resulting from the sapphire rod and the metallic oxide on the inner surfaces of the cavity are negligible compared to the contributions of the dielectric samples under test. This is ensured by the quality of the sapphire used \cite{read2023precision} and the chemical treatments performed on the cavity \cite{kudra2020high}. In addition, the participation ratios of the sapphire rod for the target eigenmodes are smaller than those relative to the dielectrics, further strengthening the setup's ability to isolate the samples' contribution (see \emph{Supplementary Note~1}). The resulting participation ratios and dielectric loss tangents in the single-photon regime are summarized in Table~\ref{tab:DielLossTangSinglePhoton}.

\input{Tables/Dielectric_loss_tangent_different_modes}

\subsection*{Temperature and power sweeps}

To fully characterize the dielectric behavior of the samples, we investigate the power and temperature dependence of the dielectric losses. For power sweep measurements, the input signal is varied from the maximum VNA power, i.e., \hspace{3pt}20\hspace{3pt}dBm, to a minimum magnitude that would be equivalent to an average cavity photon number of $\bar{n}_{ph} = 1$ to probe the material's behavior in the single-photon regime. The average photon number present in the cavity is obtained through the relation for reflection configuration \cite{Bruno2014Reducing}:

\begin{equation}
    \label{eq:AvgPhNumb}
    \bar{n}_{ph} = \frac{4}{h\nu_{0}^{2}}\frac{Q_{l}^{2}}{Q_{ext}}P_{in},
\end{equation}
where $P_{in}$ is the input power in Watts at the cavity port. The power sweeps are performed at the base temperature of the fridge, circa 10\hspace{3pt}mK. The input magnitude at the cavity port is determined by performing a power loss characterization of the wiring setup with a spectrum analyzer at frequencies detuned from the resonant modes of interest by more than ten times their linewidth. 

Temperature sweeps are conducted by adjusting the fridge's mixing chamber temperature from 10~mK to 700~mK for the AlN sample and from 10~mK to 1.5~K for the LiNbO$_{3}$ samples. After reaching the target temperature, proper thermalization of the cavity and sample is confirmed by a ruthenium oxide thermometer mounted on the cavity, as well as by the VNA $S_{11}$ trace stability.

\subsection*{Methods for XPS and ToF-SIMS analysis}
XPS data taken from surfaces of AlN and LiNbO$_{3}$ bulk crystals were collected using a SPECS FlexMod-FlexPS using a spot size of 2~mm and data analysis was performed using CasaXPS. The background signal arising from inelastically scattered electrons was removed using a Shirley baseline model, and the peaks were fit using a Gaussian/Lorentzian product. All spectra were calibrated to the C 1s peak at 285 eV. ToF-SIMS measurements were taken from the crystals using a dual beam time-of-flight secondary ion mass spectrometry (IONTOF 5) to analyze the concentration and depth distribution of impurities. Secondary ion measurements were performed using a liquid bismuth ion beam (Bi$^+$). A cesium ion gun with an energy of 500 eV was used for sputtering the surface for depth profile measurements to enhance the detection of anions \cite{bose2015study}. A 100 \textmu m by 100 \textmu m analysis area and 400 \textmu m by 400 \textmu m sputter area were used for all measurements. Because depth profiling is destructive, the analysis was performed on separate AlN and LiNbO$_{3}$ samples taken from the same lot of crystals as those inserted into the cavities.

%% file: Tables/Dielectric_loss_tangent_different_modes.tex
\setlength{\tabcolsep}{4pt} 
\begin{center}
\begin{table}[htbp!]
    \centering
    \caption{Dielectric loss tangent at base temperature in the single-photon power regime for all resonant modes of the samples.}
    \begin{tabular*}{0.78\textwidth}{@{}lcccc@{}}
    \toprule
    Sample & Mode & $Q_{0}\left(\bar{n}_{ph}\rightarrow0\right)$ $\left(\times 10^{4}\right)$ & $p \ [\%]$ & $\tan\delta\left(\bar{n}_{ph}\rightarrow0\right) \left(\times 10^{-5}\right)$ \\
    \midrule
        \multirow{2}{*}{AlN} & TM$_{010}$ &  8.29 $\pm$ 0.02 & 48.9 $\pm$ 0.1  & 2.47 $\pm$ 0.04 \\
        & TE$_{011}$ & 7.0 $\pm$ 0.4 & 93.3 $\pm$ 0.1 & 1.5 $\pm$ 0.6 \\
        \multirow{2}{*}{LiNbO$_{3}$} & TM$_{010}$ &  11.1 $\pm$ 0.1 & 44.9 $\pm$ 0.1 & 2.01 $\pm$ 0.01 \\
        & TE$_{011}$ & 4.68 $\pm$ 0.04 & 94.1 $\pm$ 0.1& 2.27 $\pm$ 0.09\\
    \bottomrule
    \end{tabular*}\label{tab:DielLossTangSinglePhoton}
\end{table}
\end{center}

%% file: conclusions.tex
We performed a detailed characterization of the microwave properties of two widely used electro-optic materials, AlN and LiNbO$_{3}$. The study is motivated by the open question of whether material-specific spurious loss and coupling channels, such as electromechanical/piezoelectric-related dissipation and anisotropy-driven effects, that are known to be problematic in thin-film and micrometer-scale implementations, persist in bulk crystals at millikelvin temperatures. Using a robust circle-fit routine~\cite{probst2015efficient}, we extracted the single-photon internal quality factor in the millikelvin regime for two excitation directions of the anisotropic dielectric tensor.

We find that the intrinsic microwave quality factors of both crystals fall within the range required for high-efficiency microwave–optical quantum transduction~\cite{wang2022high-efficiency}. These results support the premise that macroscopic electro-optic crystals integrated in hybrid devices can achieve substantially lower microwave losses compared to lithographically fabricated devices, where interface participation and parasitic loss channels often limit the microwave quality factor. Since microwave–optical conversion efficiency is directly limited by the microwave quality factor, improving this parameter is essential for achieving high-efficiency transduction in hybrid quantum architectures.

We further characterized the intrinsic quality factor and the loss tangent as a function of microwave power. At low power, the loss mechanisms are well described by a standard TLS loss model~\cite{gao2008thephysics,gao2008equivalence}. TLS saturation occurs at comparatively low intracavity photon number, consistent with a loss mechanism dominated by a surface/near-surface TLS population rather than bulk-distributed defects. At higher power, we observe a distinct logarithmic dependence of $Q_0$ on photon number (linear in applied power in dBm) that appears only for transverse-magnetic resonances. This mode-selective behavior is consistent with a regime of strongly interacting TLS~\cite{Faoro2012Internal,Burnett2016Analysis}. Our study also excluded the presence of spurious piezoelectric losses in the analyzed temperature and power regimes. Complementary XPS and ToF-SIMS analyses corroborated the interpretation that losses are dominated by near-surface defects. In AlN we identify aluminum oxide and residual metallic aluminum bonding, while in LiNbO${_3}$ the presence of Nb$^{4+}$ indicates oxygen vacancies. These impurities and defects provide a microscopic origin for the observed TLS phenomenology and suggest that improvements in surface preparation, handling, and encapsulation could be a viable path to further reducing microwave loss. 

Overall, these results clarify which dissipation channels limit macroscopic electro-optic crystals in the quantum regime and provide quantitative guidance for designing hybrid SRF-cavity and macroscopic electro-optic crystal-based quantum transducers.

%% file: Tables/Line_characterization.tex
\begin{center}
\begin{table}[htbp!]
    \centering
    \caption{Input line attenuation from base-temperature round-trip measurements (probe power -10 dBm).}
    \begin{tabular}{@{}lccc@{}}
    \toprule
    Material & Mode & $\nu_{p}$ [GHz] & $P_{\mathrm{l}}$ [dBm] \\
    \midrule
         \multirow{2}{*}{AlN}& TM$_{010}$ & 9.560 & -60 \\
         & TE$_{011}$ & 15.500 & -90 \\
         \multirow{2}{*}{LiNbO$_{3}$}& TM$_{010}$ & 7.650 & -60 \\
         & TE$_{011}$ & 7.000 & -60 \\
    \bottomrule
    \end{tabular}
    \label{tab:LineChar}
\end{table}
\end{center}

%% file: Tables/Power_sweep_regression_results_TM010_TE011_full.tex
\begin{sidewaystable}
\centering
\begin{threeparttable}
    \centering
    \caption{Piecewise fit results of the TM$_{010}$ mode internal quality factor versus power data series of the two materials.}
    \begin{tabular*}{0.9\textwidth}{@{}lccccccc@{}}
    \toprule
    Sample & $Q_{\mathrm{TLS,0}} \left(\times 10^5\right)$ & $P_{c}$ [dBm] & $\beta$ & $Q_{\mathrm{other}} \left(\times 10^5\right)$ & $Q_{0}\left(\bar{n}_{ph}\rightarrow0\right) \left(\times 10^5\right)$ & $Q_{\log} \left(\times 10^3\right)$ & $P_{c,\log}$ [dBm]  \\
    \midrule
        {AlN\tnote{1}} & 3.59 $\pm$ 0.04 & -70.3 $\pm$ 0.1  & 0.77 $\pm$ 0.02 & 1.08 $\pm$ 0.01 &  0.829 $\pm$ 0.02 & 2.39 $\pm$ 0.05 & -50 $\pm$ 1 \\
        {LiNbO$_{3}$\tnote{2}} & 8.47 $\pm$ 0.04 & -93.9 $\pm$ 0.3 & 0.60 $\pm$ 0.02 & 1.27 $\pm$ 0.03 & 1.11 $\pm$ 0.01 & 0.98 $\pm$ 0.01 & -60 $\pm$ 1 \\
    \bottomrule
    \end{tabular*}
    \begin{tablenotes}
        \item[1] {Resonance frequency $\nu_{0}=$ 9.5661021(2) GHz }
        \item[2] {Resonance frequency $\nu_{0}=$ 7.6750283(4) GHz }
    \end{tablenotes}
    \label{tab:PwrSweepRegrResultsTM010}
    
\end{threeparttable}

\bigskip\bigskip\bigskip\bigskip\bigskip\bigskip\bigskip\bigskip\bigskip\bigskip

\begin{threeparttable}
    \caption{Fit results of the TE$_{011}$ mode internal quality factor versus power data series of the two materials for different DR runs.}
    \begin{tabular*}{0.69\textwidth}{@{}lccccc@{}}
    \toprule
    Sample & $Q_{\mathrm{TLS,0}} \left(\times 10^5\right)$ & $P_{c}$ [dBm] & $\beta$ & $Q_{\mathrm{other}} \left(\times 10^4\right)$ & $Q_{0}\left(\bar{n}_{ph}\rightarrow0\right) \left(\times 10^4\right)$ \\
    \midrule
        AlN\tnote{1} & 7.2 $\pm$ 0.6 & -75 $\pm$ 1 & 1.4 $\pm$ 0.5 & 7.8 $\pm$ 0.3 & 7.0 $\pm$ 0.4\\
        LiNbO$_{3}$\tnote{2} & 3.64 $\pm$ 0.03 & -84.9 $\pm$ 0.2 & 1.05 $\pm$ 0.03 & 5.38 $\pm$ 0.04 & 4.68 $\pm$ 0.04 \\
    \bottomrule
    \end{tabular*}
    \begin{tablenotes}
        \item [1]{Resonance frequency $\nu_{0}=$ 15.508453(4) GHz}
        \item [2]{Resonance frequency $\nu_{0}=$ 7.020667(1) GHz}
    \end{tablenotes}
    \label{tab:PwrSweepRegrResultsTE011}
\end{threeparttable}
\end{sidewaystable}